\documentclass[a4paper,twocolumn,11pt,unpublished]{quantumarticle}
\pdfoutput=1
\usepackage[utf8]{inputenc}
\usepackage[english]{babel}
\usepackage[T1]{fontenc}
\usepackage{amsmath,amssymb,amsthm}
\usepackage{hyperref}

\usepackage{tikz}
\usepackage{lipsum}

\usepackage[numbers]{natbib}

\usepackage{comment}
\usepackage{amsthm}
\usepackage{mathtools}

\usepackage{braket}

\usepackage{subcaption}
\usepackage{graphicx}

\newtheorem{theorem}{\bf Theorem}[section]

\usepackage{bbm}

\newcommand\rmd{\mathrm{d}}
\newcommand\rme{\mathrm{e}}

\newcommand\rmA{\mathrm{A}}
\newcommand\rmB{\mathrm{B}}

\newcommand\rmS{\mathrm{S}}

\newcommand\bbr{\mathbb{R}}

\newcommand\calA{\mathcal{A}}

\newcommand\calE{\mathcal{E}}
\newcommand\calF{\mathcal{F}}
\newcommand\calG{\mathcal{G}}
\newcommand\calH{\mathcal{H}}

\newcommand\calL{\mathcal{L}}
\newcommand\calM{\mathcal{M}}

\newcommand\calO{\mathcal{O}}

\newcommand\calR{\mathcal{R}}

\newcommand\calU{\mathcal{U}}
\newcommand\calV{\mathcal{V}}

\newcommand\rmtr{\mathrm{tr}}

\begin{document}

\title{Robustness of quantum algorithms:
Worst-case fidelity bounds and implications for design}

\author{Julian Berberich}
\affiliation{University of Stuttgart, Institute for Systems Theory and Automatic Control and Center for Integrated Quantum Science and Technology (IQST), 70569 Stuttgart, Germany}
\orcid{0000-0001-6366-6238}
\email{julian.berberich@ist.uni-stuttgart.de}

\author{Tobias Fellner}
\affiliation{University of Stuttgart, Institute for Computational Physics, 70569 Stuttgart, Germany}
\orcid{0009-0001-2319-5635}

\author{Robert L. Kosut}
\affiliation{SC Solutions, San Jose, CA, USA, Quantum Elements, Inc., Thousand Oaks, CA, USA, and Princeton University, Princeton, NJ, USA}
\orcid{0000-0003-1206-0206}

\author{Christian Holm}
\affiliation{University of Stuttgart, Institute for Computational Physics, 70569 Stuttgart, Germany}
\orcid{0000-0003-2739-310X}

\maketitle

\begin{abstract}
Errors occurring on noisy hardware pose a key challenge to reliable quantum computing.
Existing techniques such as error correction, mitigation, or suppression typically separate the error handling from the algorithm analysis and design.
In this paper, we develop an alternative, algorithm-centered framework for understanding and improving the robustness against errors.
For a given quantum algorithm and error model, we derive worst-case fidelity bounds which can be efficiently computed to certify the robustness.
We consider general error models including coherent and (Markovian) incoherent errors and allowing for set-based error descriptions to address uncertainty or time-dependence in the errors.
Our results give rise to guidelines for robust algorithm design and compilation by optimizing our theoretical robustness measure.
We demonstrate the practicality of the framework with numerical results on algorithm analysis and robust optimization, including the robustness analysis of a 50-qubit modular adder circuit.
\end{abstract}

\section{Introduction}

Errors pose a crucial obstacle for realizing quantum algorithms on current quantum computing hardware.
Key approaches for handling errors include quantum error correction~\cite{nielsen2011quantum}, quantum error mitigation~\cite{cai2023quantum}, and quantum error suppression techniques such as dynamical decoupling or decoherence-free subspaces~\cite{lidar2014review}.
However, they cannot completely eliminate errors on noisy intermediate-scale quantum (NISQ) devices~\cite{preskill2018quantum}.
Even when quantum error correction will be able to substantially reduce errors on early fault-tolerant~\cite{katabarwa2024early} or Megaquop~\cite{preskill2025beyond} machines, the effect of errors cannot be expected to vanish completely in the near future.
Instead, by combining different error handling techniques, noise needs to be tackled as effectively as possible to allow for the implementation of quantum algorithms of increasing size.

In this paper, we tackle the problem of errors in quantum computing from an algorithmic viewpoint.
While existing techniques such as quantum error correction, mitigation, or suppression typically separate the error handling from the algorithm analysis and design, we develop a framework for studying inherent robustness properties of quantum algorithms.
To be precise, we derive worst-case fidelity bounds against a general class of errors, including coherent and incoherent errors which can vary over time and between different qubits.
To this end, we combine set membership uncertainty descriptions~\cite{milanese1991optimal,kosut1992set,milanese2004set}, interaction picture modeling, and robust optimization~\cite{bental2009robust}.
Our results yield an efficiently computable robustness measure showing that the robustness of an algorithm depends on the precise form of the error.
Thus, a given algorithm might be more or less robust against one type of error compared to another type.
Conversely, one algorithm can be more or less robust against a specific error than another algorithm, even when both have similar depth and gate count and they implement the same unitary in the absence of noise.
Beyond explicitly quantifying the robustness of an algorithm, our findings can be used to design quantum algorithms or find compiled circuits which are inherently more robust and can, therefore, be implemented more reliably on noisy hardware.

Various existing works highlight the importance of taking hardware imperfections into account for the design and compilation of quantum algorithms~\cite{chong2017programming}.
For example, this has motivated specific algorithms for NISQ devices~\cite{preskill2018quantum,bharti2022noisy} such as variational quantum algorithms (VQAs)~\cite{cerezo2021variational}, as well as tailored algorithms for early fault-tolerant quantum computers~\cite{katabarwa2024early} such as amplitude estimation~\cite{tanaka2021amplitude,wang2021minimizing}. 
Similarly, noise-aware compilation methods~\cite{wagner2024optimized,seif2024suppressing} aim to find a robust compilation of an algorithm for a given error class.
However, they typically assume precise knowledge of the error, which can only be obtained via error characterization techniques when errors are constant over time, thus excluding time-varying errors~\cite{soare2014experimental,ji2022calibration}.
Further, a large body of existing compilation methods focuses on reducing the depth, gate count, or the number of entangling gates~\cite{maslov2008quantum,arabzadeh2010rule,nam2018automated,amy2019controlled,lee2019hybrid,duncan2020graph,foesel2021quantum,nagarajan2021quantum}.
The robustness analysis of quantum error correction schemes against specific errors via tailored simulation techniques has also received substantial attention~\cite{bravyi2018correcting,marton2023coherent}.
On the other hand, there are only few results which rigorously quantify the robustness of a generic quantum algorithm against general error classes.

In~\cite{berberich2024robustness}, results in this direction were obtained in the form of worst-case fidelity bounds for the specific class of coherent control errors, which are multiplicative Hamiltonian errors that can be caused, e.g., by miscalibration.
Analogous results for coherent control errors in quantum annealing were obtained in~\cite{funcke2024robustness}.
Further robustness analysis results include~\cite{abad2022universal,abad2025impact}, which study the average fidelity of individual gates in the presence of incoherent errors.
Moreover, the recent paper~\cite{garcia2025resilience} quantifies the average fidelity of an algorithm subject to noise using tools from quantum information geometry.
They apply their theoretical results to derive resilience-runtime tradeoffs, showing, e.g., that longer algorithms can be more robust than shorter ones.
However,~\cite{garcia2025resilience} handles incoherent errors by averaging over coherent errors, which excludes relaxation errors such as amplitude damping.
Further, the recent paper~\cite{sannia2025uncovering} employs metastability for scalable robustness analysis of quantum algorithms, assuming Pauli noise and without deriving explicit fidelity bounds.
Finally, related robustness questions were studied in the context of quantum control~\cite{kosut2013robust,kosut2022robust}.

In the context of these works, our main contributions are:
\begin{enumerate}
    \item the derivation of explicitly computable worst-case fidelity bounds as required in standard threshold theorems for fault-tolerant quantum computing~\cite{nielsen2011quantum,kitaev1997quantum,aharonov1997fault,aharonov2008fault};

    \item a general and scalable robustness analysis framework which can handle arbitrary coherent and (stationary or time-dependent Markovian) incoherent errors;

    \item quantitative insights into inherent robustness properties and fundamental limitations of quantum algorithms;

    \item a systematic approach for robust algorithm design and compilation, including an application to finding composite pulses~\cite{levitt1986composite,jones2003robust,brown2004arbitrarily,merrill2014progress} with specified robustness properties.
\end{enumerate}

The remainder of the paper is structured as follows.
In Section~\ref{sec:robustness_quantum_algorithms}, we derive worst-case fidelity bounds for quantum algorithms subject to coherent errors.
We develop explicit procedures for efficiently and accurately computing the bounds for a given algorithm and error model, and we discuss implications of our results for analysis and design.
Next, in Section~\ref{sec:robustness_quantum_operations}, we provide analogous results for more general quantum algorithms (including invertible quantum operations) and error classes (including incoherent errors).
In Section~\ref{sec:numerical_results}, we apply our theoretical findings in simulation to analyze the robustness of given quantum algorithms (composite pulses, the quantum Fourier transform, and a $50$-qubit modular adder circuit) and to design or compile quantum algorithms with specified robustness properties.
Finally, in Section~\ref{sec:conclusion}, we conclude the paper.

\section{Robustness of quantum algorithms against coherent errors}\label{sec:robustness_quantum_algorithms}

In this section, we analyze the robustness of quantum algorithms against coherent errors. 
Coherent errors are errors that can be described by unitary operations, and they pose serious challenges for reliably implementing quantum algorithms~\cite{gottesman2019maximally,subasi2021impact,figueroa2024estimating,ott2024error}.
After introducing the problem setup (Section~\ref{subsec:algorithms_problem_setup}), we present the main result of this section: a worst-case fidelity bound (Section~\ref{subsec:algorithms_fidelity_bound}).
In Section~\ref{subsec:algorithms_discussion}, we discuss the theoretical results, and we highlight their implications for quantum algorithm design in Section~\ref{subsec:algorithms_design}.

\subsection{Problem setup}\label{subsec:algorithms_problem_setup}
Consider an ideal quantum algorithm
\begin{align}\label{eq:algo_nominal}
    \bar{U}=\bar{U}_{N}\cdots\bar{U}_1
\end{align}
composed of unitary operators $\bar{U}_j=e^{-i\bar{H}_j}$ for some Hermitian generators $\bar{H}_j=\bar{H}_j^\dagger$ \footnote{Ideal quantities are denoted by an overbar.}.
In general, the operators $\bar{U}_j$ can act on $n$ qubits, although they typically only affect a small subset of qubits non-trivially.
For example, $\bar{U}_j$ can represent a single-qubit or two-qubit gate, or multiple such gates acting on disjoint subsets of qubits.

In this section, we analyze the robustness of a generic quantum algorithm of the form~\eqref{eq:algo_nominal} against coherent errors.
To this end, we define the perturbed algorithm 
\begin{align}\label{eq:algo_perturbed}
    U=U_{N}\cdots U_1
\end{align}
with noisy gates
\begin{align}\label{eq:algo_perturbed_Uj}
    U_j=\bar{U}_jU_{\rme,j},
\end{align}
where $U_{\rme,j}$ is an error unitary representing a coherent error.
The errors $U_{\rme,j}$ may differ for each gate $U_j$ and their exact description is unknown, but it is known that they lie in the uncertainty set $\calU_{\rme}$ defined as
\begin{align}\label{eq:uncertainty}
    \Big\{ U_{\rme}=\{U_{\rme,j}\}_{j=1}^N\Big\lvert 
        U_{\rme,j}=e^{-iH_{\rme,j}},\>H_{\rme,j}\in\calH_{\rme,j}
        \Big\}.
\end{align}
Here, the Hermitian error generators $H_{\rme,j}=H_{\rme,j}^\dagger$ admit a description $H_{\rme,j}\in\calH_{\rme,j}$, which may include a uniform bound but can also impose additional structure.
For example, if the error $U_{\rme,j}$ is a Pauli-$Z$ rotation with unknown angle bounded by $\bar{\theta}$, then
\begin{align}\label{eq:z_error_example}
    \calH_{\rme,j}= \Big\{\frac{\theta_j}{2} Z\mid | \theta_j|\leq\bar{\theta}\Big\}.
\end{align}
Uncertainty descriptions such as~\eqref{eq:uncertainty} are commonly referred to as set membership uncertainty, and they are well-established for modeling uncertainty, e.g., in estimation and control~\cite{milanese1991optimal,kosut1992set,milanese2004set}.
Note the generality of the above error model, which includes arbitrary unitary operators which can vary between qubits and gates.
In particular, the set-based description includes independent errors which vary over time, e.g., between different executions of the algorithm, or between different hardware platforms, as long as the error bounds are fulfilled.
In practice, estimates of the set $\calH_{\rme,j}$ can be obtained, e.g., using tools for quantum process tomography~\cite{mohseni2008quantum} or error characterization~\cite{kaufmann2025characterization,tripathi2025benchmarking,crupi2025efficient,noeller2025classical}.

The main objective of this section is to study the deviation between the ideal quantum algorithm $\bar{U}$ and the perturbed algorithm $U$.
Their difference is quantified via the fidelity 
\begin{align}\label{eq:fidelity}
    F(\bar{U},U)=\Big|\frac{1}{2^n}\rmtr(\bar{U}^\dagger U)\Big|^2.
\end{align}
Since the errors are not known precisely but only the set-membership description $\calU_{\rme}$ is available, we aim to bound the worst-case fidelity over all errors $U_{\rme}\in\calU_{\rme}$.
To be precise, we consider the following robustness analysis problem:
For a given quantum algorithm $\bar{U}$ in~\eqref{eq:algo_nominal} and a given error model $\calU_{\rme}$ in~\eqref{eq:uncertainty}, we want to derive a lower bound on
\begin{align}
  F_{\mathrm{wc}}=\min_{U_{\rme}\in\calU_{\rme}}F(\bar{U},U).
\end{align}

\subsection{Worst-case fidelity bound}\label{subsec:algorithms_fidelity_bound}

To formally state the main result, we define the interaction Hamiltonian 
\begin{align}\label{eq:bar_G_j_def}
    G_j=\bar{V}_j^\dagger H_{\rme,j}\bar{V}_j
\end{align}
for $j=1,\dots,N$ with
\begin{align}\label{eq:bar_V_j_def}
  \bar{V}_{k}&=\bar{U}_{k-1}\cdots\bar{U}_1
\end{align}
for $k=2,\dots,N$ and $\bar{V}_1=I$.
Further, we introduce the averaged interaction Hamiltonian
\begin{align}\label{eq:averaged_interaction_Hamiltonian}
    G=\frac{1}{N}\sum_{j=1}^N G_j.
\end{align}
The following result bounds the worst-case fidelity depending on the size of the errors and the (averaged) interaction Hamiltonian.

\begin{theorem}\label{thm:avg_algo}
    For any $U_{\rme}\in\calU_{\rme}$, the fidelity is bounded as
\footnote{Throughout this paper, unless stated otherwise, norms are (induced) $2$-norms.}
    \begin{align}\label{eq:thm_avg_algo_aux}
      &F(\bar{U},U)\\\nonumber 
      \geq&1-\left(\frac{1}{2}\sum_{j=1}^N\Big\lVert \sum_{k=j+1}^N[G_j,G_k]\Big\rVert+N\lVert G\rVert\right)^2.
    \end{align}
    Further, suppose that, for all $H_{\rme,j}\in\calH_{\rme,j}$ and $j=1,\dots,N$,
    \begin{align}\label{eq:thm_avg_algo_ass1}
        \lVert H_{\rme,j}\rVert&\leq\delta,\\\label{eq:thm_avg_algo_ass2}
      \lVert G\rVert&\leq\gamma\delta,
    \end{align}
    with some $\delta,\gamma>0$.
    Then,
    \begin{align}\label{eq:thm_avg_algo}
      F_{\mathrm{wc}}\geq1-\delta^2 N^2\Big(\frac{N-1}{2}\delta+\gamma\Big)^2
    \end{align}
  \end{theorem}

  The proof of Theorem~\ref{thm:avg_algo} can be found in Appendix~\ref{app:proof_thm_avg_algo}.
Inequality~\eqref{eq:thm_avg_algo} shows that the worst-case infidelity is bounded in terms of the error level $\delta$, the depth $N$, and the norm of the averaged interaction Hamiltonian $\lVert G\rVert$.
Thus, the worst-case fidelity $F_{\mathrm{wc}}$ is high if the individual gate fidelity is high ($\delta$ is small) and the norm of the averaged interaction Hamiltonian $\lVert G\rVert$ is small.
Note that, by definition, $G$ depends linearly on the error which is why the bound~\eqref{eq:thm_avg_algo_ass2} is linear in $\delta$.

In Appendix~\ref{app:robustness_bounds}, we show how the bound $\gamma$ and, in particular, the worst-case fidelity bound~\eqref{eq:thm_avg_algo_aux} can be computed explicitly for a given quantum algorithm and error model.
We present different complementary approaches with varying levels of scalability and conservatism, which can be scaled to large numbers of qubits and gates via a partitioning approach explained in Appendix~\ref{app:circuit_partitioning}.
Together with Theorem~\ref{thm:avg_algo}, this yields a method for computing worst-case fidelity bounds for a given quantum algorithm $\bar{U}$ and an error description $\calU_{\rme}$.
We note that a related result to Theorem~\ref{thm:avg_algo} was obtained in~\cite{kosut2022robust} in the context of robust quantum control.

\subsection{Discussion}\label{subsec:algorithms_discussion}

Theorem~\ref{thm:avg_algo} implies that a given worst-case fidelity bound $\bar{F}_{\mathrm{wc}}$ is guaranteed if \footnote{The bound~\eqref{eq:scaling_infidelity_bound} uses $N-1<N$.}
\begin{align}\label{eq:scaling_infidelity_bound}
   1-\bar{F}_{\mathrm{wc}}\geq\delta^2 N^2\Big(\frac{\delta N}{2}+\gamma\Big)^2.
\end{align}
Note that the noise level $\delta$ and the algorithm depth $N$ enter multiplicatively, i.e., when the algorithm depth $N$ increases then the worst-case fidelity necessarily deteriorates unless the noise level $\delta$ decreases proportionally, and vice versa.
In Appendix~\ref{app:comparison}, we discuss why such a linear dependence $\delta\sim\frac{1}{N}$ can be expected for a worst-case fidelity bound.
There, we also discuss the relation of the bound in Theorem~\ref{thm:avg_algo} to the existing literature, in particular to the robustness analysis for the special class of coherent control errors from~\cite{berberich2024robustness} and to the average fidelity expression for stochastic noise models from~\cite{garcia2025resilience}.

The lower bound~\eqref{eq:scaling_infidelity_bound} has two main ingredients:
The term $\gamma$ quantifies the contribution of the algorithm to the worst-case fidelity, which can be leveraged for robust algorithm design or compilation, compare Section~\ref{subsec:algorithms_design}.
On the other hand, even when $\gamma=0$, the bound 
\begin{align}\label{eq:scaling_infidelity_bound_gamma_0}
1-\bar{F}_{\mathrm{wc}}\geq\frac{1}{4}(\delta N)^4,
\end{align}
remains.
Thus, the infidelity scales in a quartic fashion with $\delta N$.

\begin{figure}
    \begin{center}
    \includegraphics[width=0.5\textwidth]{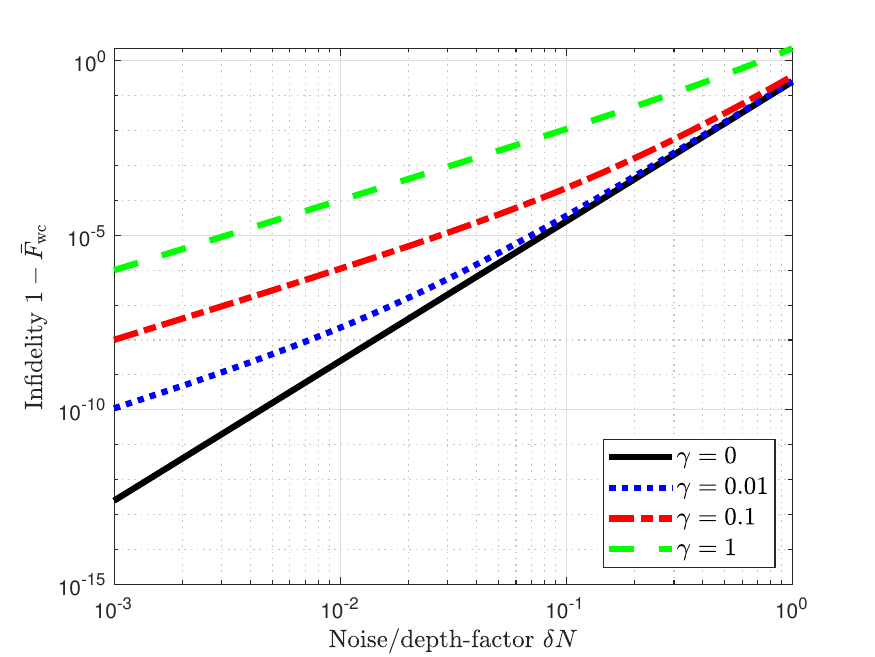}
    \end{center}
    \caption{The figure displays the infidelity bound~\eqref{eq:scaling_infidelity_bound} depending on the noise/depth-factor $\delta N$, i.e., the product of the noise level $\delta$ and the circuit depth $N$.}	\label{fig:fidelity_bound}
\end{figure}

Figure~\ref{fig:fidelity_bound} depicts the infidelity bound in~\eqref{eq:scaling_infidelity_bound} for different values of $\gamma$ and varying values of the noise/depth-factor $\delta N$, i.e., the product of the noise level $\delta$ and the algorithm depth $N$.
Let us use this plot to discuss inherent robustness properties of quantum algorithms.
First, we consider an exceptionally robust algorithm for which the average interaction Hamiltonian vanishes, i.e., $\gamma=0$, such that only the intrinsic error caused by the noise remains.
In this case,~\eqref{eq:scaling_infidelity_bound} simplifies to~\eqref{eq:scaling_infidelity_bound_gamma_0}.
Assuming a depth of $N=10^3$, a worst-case fidelity of $\bar{F}_{\mathrm{wc}}\geq0.999$ can then be guaranteed if the individual error level $\delta$ is below $2.51\cdot10^{-4}$.
Since this bound corresponds to an algorithm with perfect robustness $\gamma=0$, it provides an explicit quantification of the limitations of unitary quantum algorithms without measurements, see~\cite{kosut2022robust,kosut2025fundamental} for an analogous result in the context of robust quantum control.
At this point, for further increasing the worst-case fidelity bound, it is necessary to reduce the individual gate error level $\delta$ or the depth $N$.
Alternatively, one has to go beyond unitary quantum algorithms and include measurements and ancilla qubits, e.g., to implement quantum error correction operations.

For larger values of $\gamma$, the required value of $\delta N$ decreases further.
For example, for an algorithm with poor robustness, e.g., $\gamma=1$, achieving a worst-case fidelity bound of $\bar{F}_{\mathrm{wc}}\geq0.999$ requires a smaller error level $\delta\leq3.1\cdot10^{-5}$ when keeping the depth at $N=10^3$.
This shows that an algorithm which is designed in an inherently robust way, i.e., such that $\gamma\approx0$, admits a significant increase of the tolerable error level $\delta$.
In particular, for the fixed depth $N=10^3$ considered above, reducing $\gamma$ from $\gamma=1$ (low robustness) to $\gamma=0$ (high robustness) allows one to tolerate eight times larger error levels without reducing the guaranteed worst-case fidelity.
Due to the linear dependence $\delta\sim\frac{1}{N}$, the required error levels $\delta$ for different depths $N$ can be obtained easily, e.g., increasing $N$ by a factor of $100$ decreases the allowable error level by a factor of $100$.
In Section~\ref{subsec:algorithms_design}, we discuss how to transfer these theoretical insights into concrete methods for (optimization-based) quantum algorithm design.

It should be pointed out that Theorem~\ref{thm:avg_algo} only provides a lower bound on the worst-case fidelity and, therefore, the actual worst-case fidelity obtained by the algorithm can be larger.
In the numerical simulations below, we show that the bound is a good qualitative indication of robustness, i.e., it directly correlates with empirical fidelity estimates.
Further reducing conservatism to obtain more accurate robustness measures is an interesting direction for future research.

Next, let us discuss the influence of the number of qubits on the bound in Theorem~\ref{thm:avg_algo}.
To this end, note that multiple errors acting on $\ell$ of the qubits can be addressed via $H_{\rme,j}=\sum_{k=1}^{\ell}\theta_{j,k}B_{j,k}$ with $\lVert B_{j,k}\rVert\leq1$ and unknown but bounded error angles $\theta_j=(\theta_{j,1},\dots,\theta_{j,k})$.
In that case,~\eqref{eq:thm_avg_algo_ass1} is guaranteed as long as $\lVert\theta_j\rVert\leq\delta$, $j=1,\dots,N$.
Thus, for errors $\theta_j$ which are bounded in the $2$-norm, the worst-case fidelity bound~\eqref{eq:thm_avg_algo} is independent of the number of qubits.
Alternatively, an infinity-norm bound on $\theta_j$ such as $|\theta_{j,k}|\leq\tilde{\delta}$ for $j=1,\dots,N$, $k=1,\dots,\ell$ implies that~\eqref{eq:thm_avg_algo_ass1} holds with $\delta=\tilde{\delta}\ell$.
Hence, for errors $\theta_j$ bounded in the $\infty$-norm, the worst-case fidelity guaranteed via Theorem~\ref{thm:avg_algo} scales as $\mathcal{O}(\ell^4)$ when $\ell$ of the qubits are affected by individual errors.

Finally, while this section focuses on unitary quantum algorithms, it can also include non-unitary effects via bipartite systems with system-bath interaction.
To be precise, suppose that the ideal unitary gate $\bar{U}_j$ is given by
\begin{align}
    \bar{U}_j=\bar{U}_{\rmS,j}\otimes I_{\rmB},
\end{align}
where $\bar{U}_{\rmS,j}$ acts only on the main system and the identity $I_{\rmB}$ acts on the environment.
Further, suppose that the error Hamiltonian defining the unitary error $U_{\rme,j}=e^{-iH_{\rme,j}}$ is equal to
\begin{align}
    H_{\rme,j}= H_{\rme\rmS,j}\otimes I_{\rmB}+I_{\rmS}\otimes H_{\rme\rmB,j}+H_{\rme\rmS\rmB,j}
\end{align}
with the identity $I_{\rmS}$ acting on the main system and error Hamiltonians for the system $H_{\rme\rmS,j}$, bath $H_{\rme\rmB,j}$, and system-bath interaction $H_{\rme\rmS\rmB,j}$.
All results in this section equally apply to such bipartite systems which allows us to model incoherent effects as well.
However, computing the fidelity bound in Theorem~\ref{thm:avg_algo} requires us to keep track of the bath state without allowing for reset, which limits the applicability for practical noise models.
Hence, in Section~\ref{sec:robustness_quantum_operations}, we present an alternative and more flexible approach for handling decoherence by modeling the ideal algorithm and the errors as vectorized quantum operations.

\subsection{Implications for quantum algorithm design}\label{subsec:algorithms_design}

In this section, we discuss the implications of Theorem~\ref{thm:avg_algo} on quantum algorithm design and compilation.
The result shows that the norm of $G$ can be used as a simple quantification of robustness of an algorithm against a given error model.
In particular, according to~\eqref{eq:thm_avg_algo}, smaller values of the bound $\gamma$ in~\eqref{eq:thm_avg_algo_ass2} lead to a larger guaranteed worst-case fidelity.
Hence, circuits with smaller $\gamma$ admit a higher degree of guaranteed robustness against the specified error model $\calU_{\rme}$ and can, therefore, be implemented more reliably on noisy quantum hardware.
This motivates taking the minimization of $\gamma$ as an additional objective into account during algorithm design or when mapping a given algorithm to the hardware via circuit compilation.

In the following, we illustrate the basic idea via optimization-based algorithm design.
To this end, suppose that (some of) the ideal quantum gates $\bar{U}_j$ in the algorithm $\bar{U}$ are parameterized, i.e., the overall algorithm takes the form
\begin{align}\label{eq:PQC}
    \bar{U}(\eta)=\bar{U}_N(\eta_N)\cdots\bar{U}_1(\eta_1).
\end{align}
For example, $\bar{U}_j(\eta_j)=e^{-i\eta_j\bar{H}_j}$ with real-valued parameters $\eta_j\in\bbr$ and Hermitian generators $\bar{H}_j$ as in VQAs~\cite{cerezo2021variational}.
For parameterized quantum circuits as in~\eqref{eq:PQC}, a common objective is to adapt the parameters $\eta_j$ in order to minimize a certain cost function $f(\eta)$, which can encode ground state search~\cite{peruzzo2014variational}, combinatorial optimization~\cite{farhi2014quantum}, but also circuit compilation~\cite{richter2024quantum}.

Consider now the following multi-objective optimization problem:
Minimizing the cost function $f(\eta)$ as well as maximizing the resilience of the algorithm against errors.
For parameterized quantum circuits as in~\eqref{eq:PQC}, the corresponding averaged interaction Hamiltonian $G(\eta)$ in~\eqref{eq:averaged_interaction_Hamiltonian} as well as its bound $\gamma(\eta)$ in~\eqref{eq:thm_avg_algo_ass2} both depend on the parameters $\eta$.
Thus, the above-mentioned multi-objective problem can be formulated as minimizing both the cost $f(\eta)$ as well as the robustness measure $\gamma(\eta)$.
In practice, this multi-objective optimization problem can be handled either by minimizing a weighted sum of the two terms
\begin{align}
    \min_{\eta}f(\eta)+\lambda\gamma(\eta)
\end{align}
with some weight $\lambda>0$, or by minimizing one of them and enforcing a hard constraint on the other one.
For example, the optimization problem
\begin{align}
    \min_{\gamma(\eta)\leq c}f(\eta)
\end{align}
minimizes the cost $f(\eta)$ while guaranteeing a specified robustness that is quantified via the parameter $c>0$. 
In Section~\ref{subsec:numerical_results_composite_pulses}, we apply this idea for designing composite pulse sequences with specified robustness properties.
One can also define multiple robustness measures $\gamma_l(\eta)$, each of which corresponds to a different error class (e.g., Pauli X/Y/Z rotations, coherent control errors), and minimize a weighted sum 
\begin{align}
    \min_{\eta}f(\eta)+\sum_{l}\lambda_l\gamma_l(\eta),
\end{align}
where the parameters $\lambda_l>0$ can be tuned to trade off the different objectives.

Note that the above algorithm-centered approach can be directly combined with alternative error handling techniques such as error correction, mitigation, or suppression.
In particular, by reducing the influence of errors on the algorithm, the application of error correction, mitigation, or suppression methods is simplified, thus improving the overall reliability of the quantum algorithm when implemented on noisy hardware. 

\section{Robustness of quantum operations against coherent and incoherent errors}\label{sec:robustness_quantum_operations}
In this section, we generalize the framework from Section~\ref{sec:robustness_quantum_algorithms} by modeling circuit elements as well as errors as quantum operations.
This allows us to consider not only more general (possibly non-unitary) circuit operations but also more general classes of errors including both coherent and (Markovian) incoherent errors.
After introducing the problem setup in Section~\ref{subsec:quantum_operations_problem_setup}, we present our main result in the form of a worst-case fidelity bound (Section~\ref{subsec:quantum_operations_fidelity_bound}).

\subsection{Problem setup}\label{subsec:quantum_operations_problem_setup}

In this section, we assume that the ideal quantum circuit is described by a quantum operation
\begin{align}
    \bar{\calE}=\bar{\calE}_N\circ\dots\circ\bar{\calE}_1
\end{align}
with individual quantum operations
\begin{align}\label{eq:E_bar_j_def}
    \bar{\calE}_j(\rho)=\sum_k\bar{E}_{j,k}\rho\bar{E}_{j,k}^\dagger    
\end{align}
acting on $n$ qubits or subsets thereof.
We assume that the individual operations $\bar{\calE}_j$ admit a mathematical inverse (which need not correspond to a physically realizable quantum channel).
This is the case, e.g., when $\bar{\calE}_j$ corresponds to a unitary quantum gate.
The ideal quantum operations~\eqref{eq:E_bar_j_def} are affected by errors, leading to the perturbed quantum operation 
\begin{align}\label{eq:E_def}
    \calE=\calE_N\circ\dots\circ\calE_1
\end{align}
with 
\begin{align}
    \calE_j=\bar{\calE}_j\circ\calE_{\rme,j}
\end{align}
for some error operation 
\begin{align}
    \calE_{\rme,j}(\rho)&=\sum_k E_{\rme,j,k}\rho E_{\rme,j,k}^\dagger.
\end{align}
Here, the individual errors $\calE_{\rme,j}$ can correspond, e.g., to coherent or incoherent errors or to a combination thereof.
In the following derivations, we will make extensive use of vectorization operators~\cite[Section 10.2.2]{petersen2012matrix}.
In particular, we write the quantum operation~\eqref{eq:E_def} as
\begin{align}\label{eq:quantum_operation_vectorized}
    \mathrm{vec}(\calE(\rho))
    =\calA\mathrm{vec}(\rho)
\end{align}
with 
\begin{align}\label{eq:quantum_operation_vectorized_ingredients}
    \calA&=\bar{\calA}_N \calA_{\rme,N}\cdots\bar{\calA}_1 \calA_{\rme,1},\\\nonumber 
    \bar{\calA}_j&=\sum_k \bar{E}_{j,k}^*\otimes \bar{E}_{j,k},\\\nonumber 
    \calA_{\rme,j}&=\sum_k E_{\rme,j,k}^*\otimes E_{\rme,j,k}.
\end{align}
We assume that the error operators $\calA_{\rme,j}$ can be represented as matrix exponentials 
\begin{align}\label{eq:quantum_operation_error_matrix}
    \calA_{\rme,j}=e^{M_{\rme,j}}.
\end{align}
Note that this is always possible for Markovian quantum operations $\calE_{\rme,j}$ which can be described via a Lindblad master equation
\begin{align}\label{eq:quantum_operations_Lindblad_equation}
    \dot{\rho}=-i[K_j,\rho]+L_j(\rho)
\end{align}
with unitary components $K_j$ and non-unitary components
\begin{align}
    L_j(\rho)=\sum_{k=1}^p2L_{j,k}\rho L_{j,k}^\dagger-(L_{j,k}L_{j,k}^\dagger\rho+\rho L_{j,k}L_{j,k}^\dagger).
\end{align}
In this case,~\eqref{eq:quantum_operation_error_matrix} holds with \footnote{For notational simplicity, we assume time evolution for $\Delta t=1$.}
\begin{align}
    M_{\rme,j}=&-i(I_{2^n}\otimes K_j-K_j\otimes I_{2^n})\\\nonumber 
    &+\sum_k
    2(L_{j,k}^*\otimes L_{j,k})\\\nonumber 
    &\quad-(I_{2^n}\otimes L_{j,k}^\dagger L_{j,k}+(L_{j,k}^{\dagger}L_{j,k})^\top\otimes I_{2^n}).
\end{align}
As in Section~\ref{sec:robustness_quantum_algorithms}, the precise form of the quantum operation defined via~\eqref{eq:quantum_operations_Lindblad_equation} is unknown, i.e., the matrices $M_{\rme,j}$ are unknown.
Instead, we assume that an uncertainty set describing the vectorized errors $\calA_{\rme,j}$ is available, i.e., a set $\hat{\calA}_{\rme}$ of the form
\begin{align}\label{eq:uncertainty_incoherent}
    \Big\{ \calA_{\rme}=\{\calA_{\rme,j}\}_{j=1}^N\Big\lvert 
        \calA_{\rme,j}=e^{M_{\rme,j}},\>M_{\rme,j}\in\calM_{\rme,j}
        \Big\},
\end{align}
with individual uncertainty sets $\calM_{\rme,j}$.
This allows us to incorporate estimates of the error channels along with error bounds.
For example, this error model includes Pauli rotations with unknown but bounded angles that occur with a certain probability, or incoherent errors (e.g., depolarization, bit-flip, phase-flip, amplitude damping) for which the probability is not known precisely.
More generally, the above error model includes (possibly time-dependent) Markovian errors for which the parameters of the Lindblad master equation~\eqref{eq:quantum_operations_Lindblad_equation} are uncertain and/or may vary over time with the only requirement that the corresponding operator $\calA_\rme$ lies in the set $\hat{\calA}_\rme$.

In this section, we bound the worst-case difference between the noisy quantum operation $\calE$ and the nominal quantum operation $\bar{\calE}$.
Various measures for comparing two quantum operations were proposed in the literature.
In the following, we employ the minimal fidelity, but we note that analogous results can be derived for alternative distance measures~\cite{gilchrist2005distance}.
We first define the fidelity between two states $\rho$ and $\sigma$ as 
\begin{align}
    \calF_{\mathrm{state}}(\rho,\sigma)=\rmtr\left(\sqrt{\sqrt{\rho}\sigma\sqrt{\rho}}\right)^2.
\end{align}

Further, for two quantum operations $\Phi_1$ and $\Phi_2$, we define the induced minimal fidelity 
\begin{align}\label{eq:F_min_def}
    \calF_{\mathrm{min}}(\Phi_1,\Phi_2)=\min_{\rho}\calF_{\mathrm{state}}(\Phi_1(\rho),\Phi_2(\rho)).
\end{align}
In the following, we are interested in worst-case bounds on $\calF_{\mathrm{min}}$ for a given quantum operation $\bar{\calE}$ and an error model according to~\eqref{eq:quantum_operations_Lindblad_equation}.
More precisely, our aim is to find a lower bound on 
\begin{align}\label{eq:F_min_wc_def}
    \calF_{\mathrm{min},\mathrm{wc}}=\min_{\calA_{\rme}\in\hat{\calA}_{\rme}}\calF_{\mathrm{min}}(\bar{\calE},\calE).
\end{align}

\subsection{Worst-case fidelity bound}\label{subsec:quantum_operations_fidelity_bound}

We begin by defining the components of the ideal quantum operation 
\begin{align}
    \bar{\calV}_j=\bar{\calA}_{j-1}\cdots\bar{\calA}_1
\end{align}
and $\bar{\calV}_1=I$.
The following theoretical analysis will rely on the interaction operator 
\begin{align}
    \calR_N&=e^{\calG_N}\cdots e^{\calG_1}
\end{align}
for $\calG_j=\bar{\calV}_j^{-1} M_{\rme,j}\bar{\calV}_j$, where the operators $\bar{\calV}_j$ are invertible since the $\bar{\calE}_j$'s are invertible.
Further, we define the averaged interaction generator 
\begin{align}\label{eq:quantum_operations_G_def}
    \calG=\frac{1}{N}\sum_{j=1}^N\calG_j.
\end{align}
In the following result, we use these quantities to derive a bound on the worst-case fidelity $\calF_{\mathrm{min},\mathrm{wc}}$.
\begin{theorem}\label{thm:avg_operations}
For any $\calA_{\rme}\in\hat{\calA}_{\rme}$, the fidelity is bounded as
    \begin{align}\label{eq:thm_avg_operations_aux}
      &\calF_{\mathrm{min}}(\bar{\calE},\calE)\\\nonumber 
      \geq&1-2^{n}
      \left(\frac{1}{2}\sum_{j=1}^N\Big\lVert\sum_{k=j+1}^N[\calG_j,\calG_k]\Big\rVert 
      +2^{n/2}N\lVert\calG\rVert\right).
    \end{align}
    Further, suppose that, for all $M_{\rme,j}\in\calM_{\rme,j}$ and $j=1,\dots,N$,
    \begin{align}\label{eq:thm_avg_operations_ass1}
        \lVert M_{\rme,j}\rVert&\leq\delta,\\\label{eq:thm_avg_operations_ass2}
      \lVert \calG\rVert&\leq\gamma\delta,
    \end{align}
    with some $\delta,\gamma>0$.
    Moreover, if all $\bar{\calE}_j$ are unitary, then,
    \begin{align}\label{eq:thm_avg_operations}
      \calF_{\mathrm{wc,min}}\geq1-2^{n/2}\delta N 
      \left(\frac{N-1}{2}\delta+2^{n/2}\gamma\right).
    \end{align}
  \end{theorem}

The proof of Theorem~\ref{thm:avg_operations} can be found in Appendix~\ref{app:proof_thm_avg_operations}.
Theorem~\ref{thm:avg_operations} generalizes Theorem~\ref{thm:avg_algo} by allowing us to consider incoherent quantum errors, as long as they can be described via a Lindblad equation.
This includes a variety of error types such as depolarization, bit- and phase-flips, amplitude damping, dephasing, and more.
This generality comes at the cost of additional conservatism.
In particular, the bound~\eqref{eq:thm_avg_operations} explicitly involves the number of qubits $n$, and the dependence on $\delta N$ is linear rather than squared.
While this means that the obtained worst-case fidelity bounds are less accurate, the conceptual insights of the theoretical results remain, in particular for comparing the robustness of quantum algorithms and for designing algorithms with specified robustness properties.
Furthermore, the first bound~\eqref{eq:thm_avg_operations_aux} in Theorem~\ref{thm:avg_operations} is applicable even for non-unitary circuit elements, assuming they are invertible.
Computing the worst-case fidelity bound~\eqref{eq:thm_avg_operations} for a given algorithm and error model requires obtaining a bound $\gamma$, compare \eqref{eq:thm_avg_operations_ass2}.
This can be achieved in full analogy to Section~\ref{sec:robustness_quantum_algorithms}.
Hence, we omit the technical details and refer to Appendix~\ref{app:circuit_partitioning} which equally applies in the setup of the present section.
Likewise, the main conceptual discussion (Section~\ref{subsec:algorithms_discussion}) as well as the implications for design (Section~\ref{subsec:algorithms_design}) carry over to Theorem~\ref{thm:avg_operations}.

\section{Applications}\label{sec:numerical_results}

In this section, we apply our theoretical framework numerically in three robustness analysis and design problems.
In Section~\ref{subsec:numerical_results_composite_pulses}, we analyze the robustness of composite pulse sequences and design new pulses with superior robustness properties.
Next, in Section~\ref{subsec:numerical_results_circuit_transpilation}, we study the robustness of different transpilations of the quantum Fourier transform. Finally, in Section~\ref{subsec:modular_adder_circuit}, we study the robustness of a $50$-qubit modular adder circuit by applying circuit partitioning.
The code to reproduce the numerical experiments is publicly available on GitHub~\cite{code_robustness_2025}.

\subsection{Robustness of composite pulses}\label{subsec:numerical_results_composite_pulses}

In the following, we study the robustness of composite pulses against coherent control errors.
Coherent control errors correspond to over- or under-rotations, i.e., the ideal quantum algorithm
\begin{align}
    e^{-i\bar{H}_N}\cdots e^{-i\bar{H}_1}
\end{align}
from~\eqref{eq:algo_nominal} is replaced by 
\begin{align}\label{eq:application_noisy_circuit}
    e^{-i(1+\theta_N)\bar{H}_N}\cdots e^{-i(1+\theta_1)\bar{H}_1}
\end{align}
with the error parameters $\theta_j\in\bbr$.
In our notation, this means that the Hermitian error generators are equal to 
\begin{align}
    H_{\rme,j}=\theta_j\bar{H}_j.
\end{align}
Composite pulses are a well-studied technique to suppress coherent control errors by replacing individual operations with a sequence of several operations~\cite{levitt1986composite,jones2003robust,brown2004arbitrarily,merrill2014progress}.
In the following, we consider the approach from~\cite{jones2003robust} which replaces a Pauli-$X$ rotation with angle $\beta$ by the sequence of rotations
\begin{align}\label{eq:application_composite_pulse_sequence}
    (\beta/2)_0\>\>\pi_{\phi_1}\>\>2\pi_{\phi_2}\>\>\pi_{\phi_1}\>\>(\beta/2)_0
\end{align}
with $\phi_1=\mathrm{arccos}(-\frac{\beta}{4\pi})$, $\phi_2=3\phi_1$.
Here, $\alpha_\phi$ denotes a rotation by the angle $\alpha$ around the axis in the x-y-plane with angle $\phi$ to the x-axis.
It is shown in~\cite{jones2003robust} that the sequence~\eqref{eq:application_composite_pulse_sequence} significantly reduces the impact of coherent control errors, assuming that they are systematic errors, i.e., they affect all gates with equal magnitude.
In our framework, this means that the values $\theta_j$ in~\eqref{eq:application_noisy_circuit} are all equal.
The presence of systematic errors is a common assumption in the literature on composite pulses~\cite{levitt1986composite,jones2003robust,brown2004arbitrarily,merrill2014progress}.
Robustness against independent errors, for which the $\theta_j$'s are not necessarily equal, was studied in~\cite{kabytayev2014robustness}, assuming that the error variation is sufficiently slow.
Further,~\cite{kaplan2024correlation} analyzes the correlation threshold for the noise beyond which composite pulses improve the fidelity.

In the following, we exploit the flexibility of the proposed framework to analyze composite pulses for both systematic and independent errors.
To be precise, we apply Theorem~\ref{thm:avg_algo} to compare the robustness of a Pauli-$X$ rotation $R_X(\frac{\pi}{4})$ and the corresponding composite pulse sequence~\eqref{eq:application_composite_pulse_sequence} by Jones~\cite{jones2003robust} against coherent control errors.
We compute $\gamma$ via combinatorial optimization, see Appendix~\ref{app:robustness_bounds_3}.
In case of independent errors, we solve the problem~\eqref{eq:gamma_bound_combinatorial} directly and then use the bound~\eqref{eq:thm_avg_algo}.
In case of systematic errors, we exploit the systematic nature to compute tighter bounds by a) exploiting in the computation that all $\theta_j$'s are equal and b) using the inequality~\eqref{eq:thm_avg_algo_aux} instead of~\eqref{eq:thm_avg_algo}.

\begin{figure}
		\centering
        \begin{subfigure}{0.48\textwidth}
            \centering
            \includegraphics[width=\textwidth]{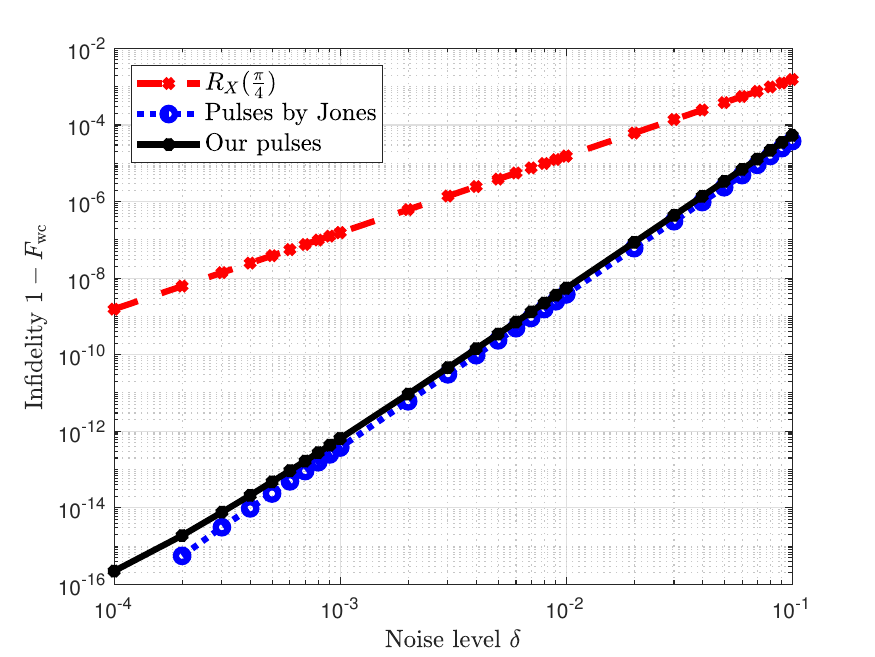}
            \caption{Systematic errors}
        \end{subfigure}
        \hfill 
        \begin{subfigure}{0.48\textwidth}
            \centering
            \includegraphics[width=\textwidth]{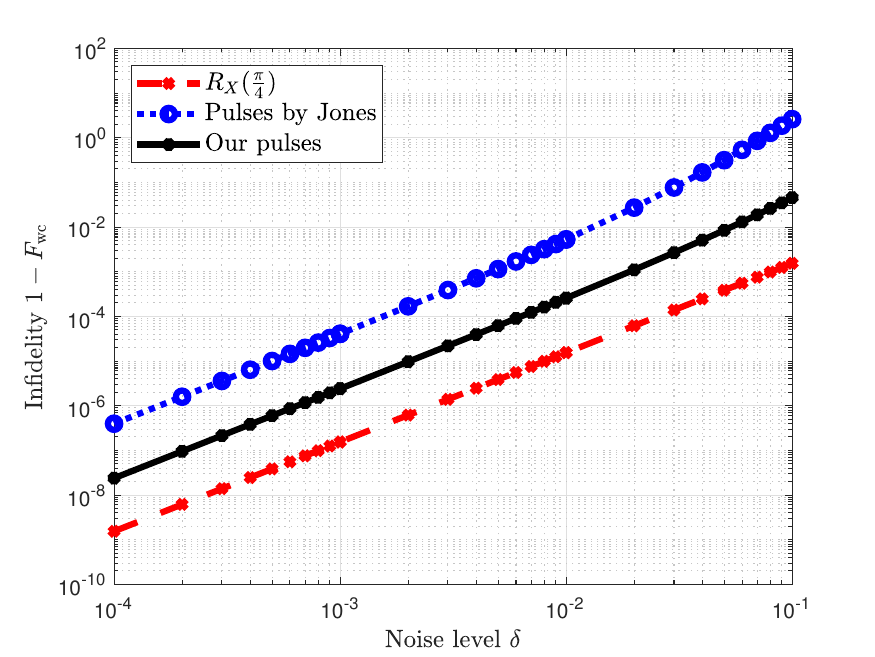}
            \caption{Independent errors}
        \end{subfigure}		
        \caption{The figures show the infidelity $1-F_{\mathrm{wc}}$ with systematic and independent errors depending on the error level $\delta$ for the Pauli-$X$ rotation $R_X(\frac{\pi}{4})$, the pulses~\eqref{eq:application_composite_pulse_sequence} by Jones~\cite{jones2003robust}, and our newly designed pulses~\eqref{eq:application_composite_pulse_sequence_ours}.
        The displayed worst-case fidelity bounds are computed based on Theorem~\ref{thm:avg_algo}.}
        \label{fig:composite_pulses_infidelity}
\end{figure}

Figure~\ref{fig:composite_pulses_infidelity} shows the worst-case infidelity for both systematic and independent coherent control errors.
Note that, indeed, the composite pulse sequence~\eqref{eq:application_composite_pulse_sequence} by Jones~\cite{jones2003robust} is substantially more robust against systematic coherent control errors than the original Pauli-$X$ rotation $R_X(\frac{\pi}{4})$, reducing the infidelity by several orders of magnitude.
This can be explained via the upper bound $\gamma$ in~\eqref{eq:thm_avg_algo_ass2} on the averaged interaction Hamiltonian.
For systematic coherent control errors, $\gamma=1$ for $R_X(\frac{\pi}{4})$ and $\gamma=0$ for the composite pulse sequence~\eqref{eq:application_composite_pulse_sequence}.
On the other hand, in the presence of independent errors, the sequence~\eqref{eq:application_composite_pulse_sequence} is significantly less robust than the original Pauli-$X$ rotation $R_X(\frac{\pi}{4})$.
Thus, the robustness benefits of composite pulses against systematic errors come along with a substantial robustness deterioration against independent errors.

Next, we leverage our theoretical analysis tools for designing new composite pulses with superior robustness properties, compare Section~\ref{subsec:algorithms_design}.
To this end, we optimize over the angles of a $5$-gate composite pulse sequence using Matlab with fmincon, based on the values from~\eqref{eq:application_composite_pulse_sequence} as initialization.
For a fixed noise level $\delta=0.05$, we maximize the worst-case fidelity for independent coherent control errors.
In the optimization, we impose as a constraint that the worst-case fidelity for systematic coherent control errors remains above $99.9995\%$, which is close to the fidelity of the composite pulses from~\cite{jones2003robust}.
The resulting sequence is given by 
\begin{align}\label{eq:application_composite_pulse_sequence_ours}
    &(0.876)_{-1.43}\>\>
    (0.834)_{0.126}\>\>
    (1.544)_{1.985}\\\nonumber 
    &(0.975)_{0.031}\>\>
    (0.808)_{-1.68}.
\end{align}
The worst-case infidelity under the newly designed pulse sequence is also shown in Figure~\ref{fig:composite_pulses_infidelity}.
The figure shows that the robustness of our pulse sequence against systematic errors is comparable to that of Jones~\cite{jones2003robust} while at the same time being significantly more robust against independent errors.
Thus, the proposed robustness framework not only allows us to compute worst-case fidelity bounds of general quantum algorithms and error models, but it can also be used to design new algorithms with specified robustness properties.
Furthermore, the framework is flexible in the sense that analogous results can be obtained for different error models, e.g., coherent Pauli-$X$/$Y$/$Z$ errors or combinations thereof.
For example, one can readily design algorithms which minimize a weighted sum of different $\gamma$'s corresponding to the averaged interaction Hamiltonians of different error classes.
In practice, this allows one to systematically tune the desired robustness properties of a quantum algoritm to a given hardware platform with error specifications.

\subsection{
Quantum circuit transpilation
}\label{subsec:numerical_results_circuit_transpilation}
We now use the proposed robustness framework to analyze the three-qubit Quantum Fourier Transform (QFT) algorithm~\cite{nielsen2011quantum}. Specifically, we study how the algorithm's robustness varies when transpiled into different elementary gate sets and subjected to various types of coherent errors. The gate sets considered are as follows:

\begin{itemize}
    \item A: $\{R_x(\pi), R_x(\pm \pi/2), R_z, iSWAP
    \}$ (used in the Rigetti Ankaa quantum processor~\cite{gateset_rigetti})
    \item B: $\{ \sqrt{X}, X, R_x, R_z, CZ, R_{zz}\}$ (used in the IBM Heron quantum processor~\cite{gateset_ibm})
    \item C: $\{PhasedXZ, R_z, SycGate, CZ, \sqrt{iSWAP}\}$ (used in the Google Sycamore quantum processor~\cite{arute_quantum_2019})
    \item D: $\{U_{1q}(\theta,\pi), U_{1q}(\theta,\pi/2), R_z, R_{zz}\}$ (used by Quantinuum~\cite{gateset_quantinuum})
\end{itemize}
The three-qubit QFT circuit is transpiled into these elementary gate sets using the Berkeley Quantum Synthesis Toolkit (BQSKit)~\cite{bqskit_software}. The original QFT circuit as well as the four transpiled circuits are provided in Appendix~\ref{app:details_numerics_qft_circuits}. 

\begin{figure}
    \centering
    \includegraphics[width=\linewidth]{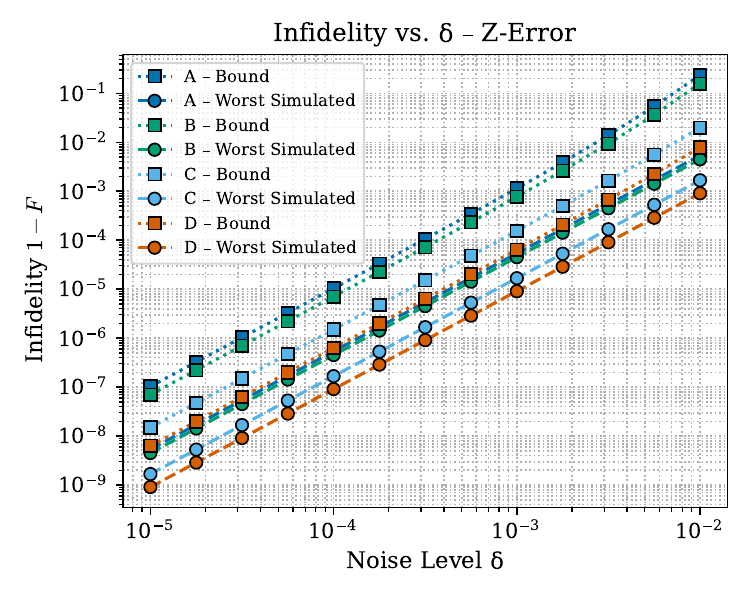}
    \caption{Infidelities of the three-qubit QFT circuit as a function of the noise level $\delta$ for different gate sets. Square markers indicate worst-case fidelity bounds computed using Theorem~\ref{thm:avg_algo}, while circular markers correspond to the worst simulated infidelity obtained from 10,000 randomly drawn noise samples. The number of gates for the transpiled circuits are: A: 59, B: 42, C: 20, D: 14.}
    \label{fig:qft_infidelity_vs_delta_z}
\end{figure}

We first consider Pauli-type coherent errors of the form
\begin{equation}\label{eq:pauli_error_qft_one_qubit}
    H_{\rme,j} = \theta_{j,1} P_m 
\end{equation}
for single qubit gates acting on qubit $m$ and 
\begin{equation}\label{eq:pauli_error_qft_two_qubit}
    H_{\rme,j} = \theta_{j,1} P_{m_2} + \theta_{j,1} P_{m_2}
\end{equation}
for two qubit gates acting on qubits $m_1$ and $m_2$. Here, $P_m$ refers to Pauli-error $P \in \{X,Y,Z\}$ acting on qubit $m$.  Additionally it holds $|\theta_{j,k}| \leq \delta$ for single qubit gates and $|\theta_{j,k}| \leq \delta$ for two qubit gates (see Appendix~\ref{app:robustness_bounds} for details on this choice).

Fidelity bounds are computed according to~\eqref{eq:thm_avg_algo}, with the parameter $\gamma$ obtained via an optimization procedure (see Appendix~\ref{app:robustness_bounds_2}). Additional details on the numerical methods used in this section are provided in Appendix~\ref{app:details_numerics_gamma_opt}. 

Figure~\ref{fig:qft_infidelity_vs_delta_z} displays the resulting worst-case infidelity bounds for $P=Z$ as a function of the noise level $\delta$ across the four gate sets.
We observe that the robustness against $Z$-errors varies with the choice of gate set. Notably, the number of layers $N$ decreases from circuit A to D, and the ordering of the worst-case fidelity bounds mirrors this trend. To validate the relevance of the bounds, we compare them with the worst-case fidelity obtained by uniformly and independently sampling $\theta_{j,i}$ across 10,000 random noise realizations. The observed correspondence between the analytical bounds and the simulation results indicates that the fidelity bound~\eqref{eq:thm_avg_algo} provides a meaningful metric for assessing the relative robustness of different transpiled circuits.

\begin{figure}
    \centering
    \includegraphics[width=\linewidth]{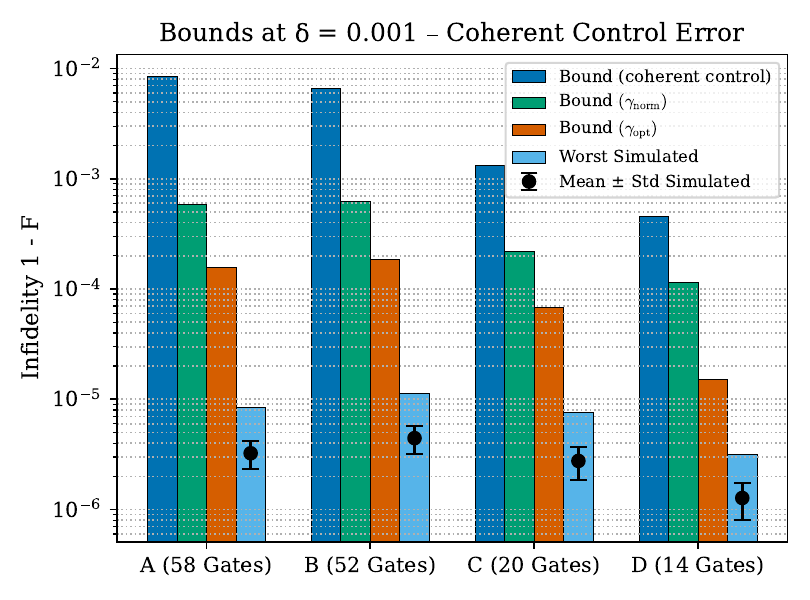}
    \caption{Infidelity bounds for coherent control errors on the transpiled QFT circuits. Theoretical bounds are computed using Theorem~\ref{thm:avg_algo} with two different strategies for obtaining $\gamma$: via an optimization problem ($\gamma_\mathrm{opt}$, Appendix~\ref{app:robustness_bounds_2}) and via a norm-based bound ($\gamma_\mathrm{norm}$, Appendix~\ref{app:robustness_bounds_4}). For comparison, the bound from~\cite[Theorem 2]{berberich2024robustness} is also shown. Additionally, we simulate 10,000 randomly sampled noise realizations, reporting both the mean and worst-case infidelity. The noise level is set to $\delta=0.001$.}
    \label{fig:qft_bounds_delta_0.001_cce}
\end{figure}

Next, we analyze the robustness of the transpiled QFT circuits under coherent control errors, as defined in~\eqref{eq:application_noisy_circuit}. Figure~\ref{fig:qft_bounds_delta_0.001_cce} compares different infidelity bounds. We first compute the bound according to Theorem~\ref{thm:avg_algo}, using two strategies to determine the parameter $\gamma$: First, $\gamma_\mathrm{opt}$ obtained via an optimization problem (Appendix~\ref{app:robustness_bounds_2}), and second, $\gamma_\mathrm{norm}$ obtained from a norm-based bound (Appendix~\ref{app:robustness_bounds_4}). By construction, $\gamma_\mathrm{opt} \leq \gamma_\mathrm{norm}$, making the infidelity bound with $\gamma_\mathrm{norm}$ more conservative. Despite this conservatism, we see that the norm-based bound is sufficient for reliably comparing the robustness of the different circuits against coherent control errors.

For reference, we also include the bound from~\cite[Theorem 2]{berberich2024robustness}. As seen in the figure, Theorem~\ref{thm:avg_algo} yields a tighter worst-case bound. A more detailed comparison between Theorem~\ref{thm:avg_algo} and~\cite[Theorem 2]{berberich2024robustness} is provided in Appendix~\ref{app:comparison_coherent_control}.

Further, we compare these theoretical bounds with simulation results obtained from $10,000$ uniformly sampled noise realizations. While the simulated worst-case fidelities are lower than the theoretical bounds, the relative robustness ranking of the circuits aligns across all considered measures. This indicates that the theoretical fidelity bounds provide a meaningful and quantitatively informative indication of algorithmic robustness under coherent control errors.
Finally, note that the simulated worst-case infidelity of circuit B is slightly larger than that of circuit A although the latter has more gates. The same relation is correctly reflected in the two bounds $\gamma_{\mathrm{opt}}$ and $\gamma_{\mathrm{norm}}$ computed based Theorem~\ref{thm:avg_algo}. On the other hand, the existing approach from~\cite{berberich2024robustness} yields a higher infidelity bound for circuit A, demonstrating the improved accuracy of the present approach.

\subsection{
Modular adder circuit
}\label{subsec:modular_adder_circuit}
One of the most promising application for quantum computing lies in factoring algorithms~\cite{shor1997polynomial}. A fundamental building block of these algorithms is the modular adder circuit~\cite{gidney_how_2025a}, which operates on two $n$-qubit registers to perform the mapping $(a,b) \xrightarrow{} (a,a+b \mod 2^n)$. 

To demonstrate the practical impact of our robustness framework on a relevant large-scale quantum computing task, we apply it to the modular adder architecture proposed in~\cite{gidney_factoring_2018} when transpiled into gate set~B. We consider a configuration with $n=25$ (totaling 50 qubits), a scale estimated to be sufficient for factoring 2048-bit RSA integers using resources expected to be available in the near future~\cite{gidney_how_2025a}.

Since a direct calculation of $\gamma$ is computationally intractable for a system of this size, we partition the circuit into smaller sub-circuits using the methodology described in Appendix~\ref{app:circuit_partitioning}. We set the maximum number of active qubits per partition to three, a value that allows for the efficient numerical calculation of $\gamma_\mathrm{opt}$ via the optimization procedure in Appendix~\ref{app:robustness_bounds_2}. The impact of the partition size on the results is further analyzed in Appendix~\ref{app:partition_scaling}.

\begin{figure}
    \centering
    \includegraphics[width=\linewidth]{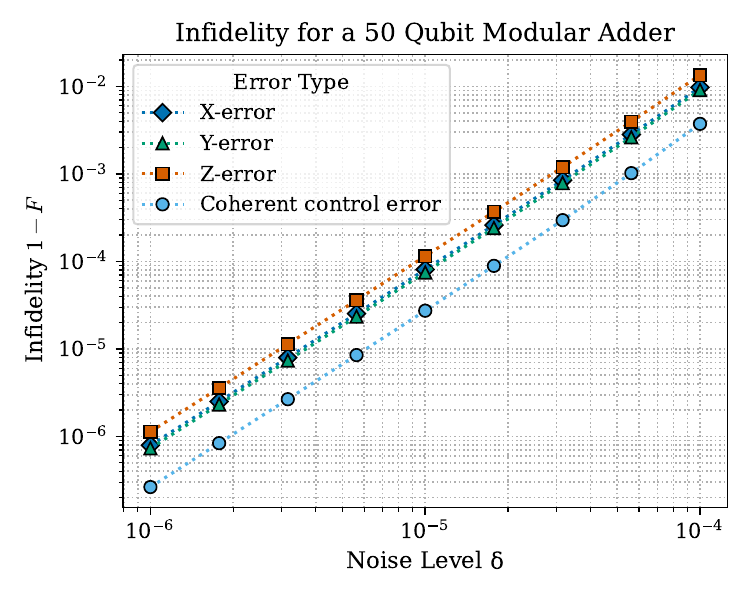}
    \caption{Infidelities of different Pauli-errors and coherent control errors over the noise level $\delta$ for a 50 qubit modular adder circuit. The infidelity bounds are computed using Theorem~\ref{thm:avg_algo} with $\gamma_\mathrm{opt}$ and partitioning into sub-circuits with at most three qubits.}
    \label{fig:adder_infidelity}
\end{figure}

By employing circuit partitioning, we are able to derive a rigorous bound on $\gamma$ even for these large-scale systems. Figure~\ref{fig:adder_infidelity} illustrates the resulting infidelity as a function of the noise level $\delta$, determined by Theorem~\ref{thm:avg_algo} for coherent Pauli-type errors (compare equations~\eqref{eq:pauli_error_qft_one_qubit} and \eqref{eq:pauli_error_qft_two_qubit}) and coherent control errors. 
Our results indicate that the robustness of the modular adder circuit against Pauli-X and Pauli-Y rotations is slightly improved in comparison to Pauli-Z rotations. Further, it exhibits substantially better robustness against coherent control errors in comparison to Pauli rotations. Thus, implementing the circuit on a hardware platform where Pauli rotation errors (especially Pauli-Z rotations) are less dominant may improve the overall reliability of the circuit.
Overall, this numerical example demonstrates that meaningful robustness guarantees can be established even for multi-qubit circuits of practically relevant size.

\section{Conclusion}\label{sec:conclusion}
We developed a framework for robustness analysis of quantum algorithms against errors.
Our results complement existing techniques such as error correction, mitigation, or suppression via an algorithm-centered perspective.
On a technical level, we derived worst-case fidelity bounds for coherent and (Markovian) incoherent errors, which may be uncertain and/or time-varying.
The derived bounds are explicitly computable and can therefore be used to study robustness of a given algorithm and error model.
We demonstrated the applicability of our framework with numerical results on algorithm analysis and robust design.

The developed framework opens the door to several promising future research directions.
Applying our results to specific design or compilation problems can yield new quantum circuits with superior robustness properties while trading off the influence on the algorithm performance, compare our results in Section~\ref{subsec:numerical_results_composite_pulses} for first results in this direction.
It will also be interesting to explore different error classes in this context such as crosstalk~\cite{murali2020software} or classes of non-Markovian errors~\cite{vega2017dynamics}.
Finally, open technical challenges include extending the results of Section~\ref{sec:robustness_quantum_operations} to more general (non-invertible) algorithm elements and to improve our theoretical analysis by deriving tighter worst-case fidelity bounds.

\section*{Acknowledgments}
The work of JB, TF, and CH was funded by Deutsche Forschungsgemeinschaft (DFG, German Research Foundation) under Germany's Excellence Strategy - EXC 2075 - 390740016. JB, TF, and CH acknowledge the support by the Stuttgart Center for Simulation Science (SimTech). 
TF and CH acknowledge support from the Deutsche Forschungsgemeinschaft (DFG, German Research Foundation) Compute Cluster Grant No. 492175459.
For RLK, this material is based upon work supported by, or in part by, the U. S. Army Research Laboratory and the U. S. Army Research Office under contract/grant number W911NF2310307.
The authors thank the International Max Planck Research School for Intelligent Systems (IMPRS-IS)
for supporting TF and CH.

\bibliographystyle{IEEEtran}
\bibliography{Literature}

\appendix

\section{Proof of Theorem~\ref{thm:avg_algo}}\label{app:proof_thm_avg_algo}

\begin{proof}
    \textbf{Part (i): Proof of~\eqref{eq:thm_avg_algo_aux}}\\
Define the interaction unitary
    \begin{align}
      R_{N}=e^{-iG_{N}}\cdots e^{-iG_{1}}.
    \end{align}
    Further, define the averaged interaction unitary 
    \begin{align}
        \bar{R}_N=e^{-iNG}.
    \end{align}
    Note that $U=\bar{U}R_N$ and, therefore, the infidelity $1-F$ is quantified by the deviation of $R_N$ from identity, i.e., 
\begin{align}\label{eq:thm_avg_algo_proof_fidelity}
    F(\bar{U},U)=\Big|\frac{1}{2^n}\rmtr(\bar{U}^\dagger\bar{U}R_N)\Big|^2=\Big|\frac{1}{2^n}\rmtr(R_N)\Big|^2.
\end{align}
    Thus, the main ingredient for proving~\eqref{eq:thm_avg_algo} will be a bound on $\lVert R_N-I\rVert$ which we establish in the following.
    To this end, we use
    \begin{align}\label{eq:thm_avg_algo_proof1}
        \lVert R_N-I\rVert\leq\lVert R_N-\bar{R}_N\rVert+\lVert\bar{R}_N-I\rVert.
    \end{align}
In order to bound the first term on the right-hand side, we use~\cite[Proposition 9]{childs2021theory} which states
\begin{align}\label{eq:thm_avg_algo_proof_trotter3}
    \lVert e^{ W_N}\cdots e^{ W_1}-e^{\sum_{j=1}^NW_j}\rVert 
    \leq&\frac{1}{2}\sum_{j=1}^N\Big\lVert \sum_{k=j+1}^N[W_j,W_k]\Big\rVert
\end{align} 
for arbitrary matrices $W_j$.
Applied to $\lVert R_N-\bar{R}_N\rVert$, we infer 
\begin{align}\label{eq:thm_avg_algo_proof_2}
    \lVert R_N-\bar{R}_N\rVert
    =&\lVert e^{-iG_N}\cdots e^{-iG_1}-e^{-i\sum_{j=1}^NG_j}\rVert\\\nonumber 
    \stackrel{\eqref{eq:thm_avg_algo_proof_trotter3}}{\leq}&
    \frac{1}{2}\sum_{j=1}^N\Big\lVert \sum_{k=j+1}^N[G_j,G_k]\Big\rVert.
\end{align}
To bound the second term on the right-hand side of~\eqref{eq:thm_avg_algo_proof1}, we define the function $f(x)=e^{-ix N G}$.
Note that $N\lVert G\rVert$ is a Lipschitz bound of $f$.
    Hence, it holds that 
    \begin{align}\label{eq:thm_avg_algo_proof_3}
        \lVert\bar{R}_N-I\rVert=\lVert f(1)-f(0)\rVert&\leq N\lVert G\rVert(1-0)=N\lVert G\rVert.
    \end{align}
    Inserting~\eqref{eq:thm_avg_algo_proof_2} and~\eqref{eq:thm_avg_algo_proof_3} into~\eqref{eq:thm_avg_algo_proof1}, we deduce 
    \begin{align}\label{eq:thm_avg_algo_proof_4}
        \lVert R_N-I\rVert\leq 
        \frac{1}{2}\sum_{j=1}^N\Big\lVert \sum_{k=j+1}^N [G_j,G_k]\Big\rVert+N\lVert G\rVert.
    \end{align}
We now use~\cite[Inequality (A14)]{kosut2022robust} which states
\begin{align}\label{eq:thm_avg_algo_proof_x}
    2\cdot2^n(1-\sqrt{F(\bar{U},U)}\leq\lVert R_N-I\rVert_F^2
\end{align}
for the Frobenius norm $\lVert\cdot\rVert_F$.
This implies 
\begin{align}\label{eq:thm_avg_algo_proof_5}
    1-\sqrt{F(\bar{U},U)}\leq &\frac{1}{2\cdot 2^n}\lVert R_N-I\rVert_F^2
    \leq\frac{1}{2}\lVert R_N-I\rVert^2.
\end{align}
Finally, using $(1-\sqrt{F(\bar{U},U)})^2\geq0$, we obtain
\begin{align}
    &1-F(\bar{U},U)\\\nonumber 
    \leq&2\left(1-\sqrt{F(\bar{U},U)}\right)\stackrel{\eqref{eq:thm_avg_algo_proof_5}}{\leq}
    \lVert R_N-I\rVert^2\\\label{eq:thm_avg_algo_proof_8}
    \stackrel{\eqref{eq:thm_avg_algo_proof_4}}{\leq}& \left(\frac{1}{2}\sum_{j=1}^N\Big\lVert \sum_{k=j+1}^N [G_j,G_k]\Big\rVert+N\lVert G\rVert\right)^2.
\end{align}
All bounds in this proof hold for arbitrary $U_{\rme}\in\calU_{\rme}$, thus proving~\eqref{eq:thm_avg_algo}.\\
\textbf{Part (ii): Proof of~\eqref{eq:thm_avg_algo}}\\
Note that 
\begin{align}\label{eq:thm_avg_algo_proof_6}
    &\frac{1}{2}\sum_{j=1}^N\Big\lVert \sum_{k=j+1}^N[G_j,G_k]\Big\rVert 
    \leq \sum_{j=1}^N\sum_{k=j+1}^N\lVert G_j\rVert\lVert G_k\rVert\\\nonumber 
    \leq&\frac{N^2-N}{2}\max_{j=1,\dots,N}\lVert G_j\rVert^2.
\end{align}
Using that $\bar{V}_j$ is unitary, we have for all $j=1,\dots,N$
\begin{align}\label{eq:thm_avg_algo_proof_7}
    \lVert G_j\rVert=\lVert H_{\rme,j}\rVert\stackrel{\eqref{eq:thm_avg_algo_ass1}}{\leq}\delta.
\end{align}
Inserting~\eqref{eq:thm_avg_algo_proof_6} and~\eqref{eq:thm_avg_algo_proof_7} into~\eqref{eq:thm_avg_algo_aux}, and using additionally~\eqref{eq:thm_avg_algo_ass2}, we obtain 
\begin{align}
    F_{\mathrm{wc}}\geq 1-\Big(\frac{N^2-N}{2}\delta^2+\gamma\delta N\Big)^2,
\end{align}
which proves~\eqref{eq:thm_avg_algo}.
\end{proof}

\section{Computation of worst-case fidelity bounds}\label{app:robustness_bounds}

In the following, we discuss how Theorem~\ref{thm:avg_algo} can be implemented computationally to obtain explicit worst-case fidelity bounds for a given quantum algorithm and error model.
Throughout this section, we assume that the $H_{\rme,j}$'s are of the form $H_{\rme,j}=\sum_{k=1}^{\ell}\theta_{j,k} B_{j,k}$ for some fixed and known matrices $B_{j,k}$ satisfying $\lVert B_{j,k}\rVert\leq1$ and with unknown but bounded error angles $\theta_j=(\theta_{j,1},\dots,\theta_{j,k})$ satisfying $\lVert\theta_j\rVert_{\infty}\leq\frac{\delta}{\ell}$, $j=1,\dots,N$.
Note that we scale the bounds on $\theta_j$ by $\frac{1}{\ell}$ to ensure that~\eqref{eq:thm_avg_algo_ass1} is fulfilled.
Alternatively, a $2$-norm bound $\lVert\theta_j\rVert\leq\delta$ can be considered.
The considered setup allows us to model multiple independent errors affecting different qubits:
For example, in a $2$-qubit system, independent Pauli-$Z$ rotation errors on each qubit can be captured by $B_{j,1}=Z\otimes I$, $B_{j,2}=I\otimes Z$.
Further, we write $G_j(\theta_j)$ and $G(\theta)$ to emphasize the dependence of $G_j$ and $G$ on the error $\theta$.

We note that the computational procedures presented in the following can be extended to more general uncertainty descriptions, following ideas from robust quantum control~\cite{kosut2022robust}.
For example, one can also handle different bounds $\delta$ and numbers $\ell$ for each error Hamiltonian $H_{\rme,j}$, different vector norms for the bound $\lVert\theta_j\rVert\leq\frac{\delta}{\ell}$, or even fully uncertain Hamiltonians $H_{\rme,j}$ without any structural knowledge of the $B_{j,k}$'s.
Moreover, while the main focus of this work lies on independent errors, for which $\theta_j$ differs for $j=1,\dots,N$, systematic errors can be analyzed as well, see Section~\ref{subsec:numerical_results_composite_pulses} for an example.

In the following, we provide four complementary approaches for computing a worst-case fidelity bound, which differ in their conservatism and computational complexity.
While the first approach tackles the bound~\eqref{eq:thm_avg_algo_aux} directly, the other three approaches employ the bound~\eqref{eq:thm_avg_algo} instead, which requires to compute a bound $\gamma$ on the norm of the averaged interaction Hamiltonian.

\subsection{Direct computation via~\eqref{eq:thm_avg_algo_aux}}
Let us reformulate the right-hand side of~\eqref{eq:thm_avg_algo_aux} as an optimization problem
\begin{align}\nonumber
    f^*=\max_{\lVert\theta\rVert_{\infty}\leq\frac{\delta}{\ell}}\Big(&\frac{1}{2}\sum_{j=1}^N\Big\lVert \sum_{k=j+1}^N[G_j(\theta_j),G_k(\theta_k)]\Big\rVert\\\label{eq:F_bound_NL}
    &+N\lVert G(\theta)\rVert\Big)^2.
\end{align}
This problem can be tackled directly using off-the-shelf solvers for nonlinear programming, e.g., interior-point or sequential quadratic programming optimization algorithms.
Based on the optimal value $f^*$, the worst-case fidelity bound can be computed as $F_{\mathrm{wc}}\geq1-f^*$.

\subsection{Reformulation of $\gamma$ as an optimization problem}\label{app:robustness_bounds_2}
We now aim to find a bound $\gamma$ such that
\begin{align}
    \lVert G(\theta)\rVert\leq\gamma\delta
\end{align}
holds for any error $U_\rme\in\calU_\rme$, compare~\eqref{eq:thm_avg_algo_ass1}.
By definition, the smallest possible worst-case bound $\gamma$ is equal to
\begin{align}\label{eq:gamma_bound_NL}
    \gamma=\max_{\lVert\theta\rVert_{\infty}\leq\frac{1}{\ell}}\lVert G(\theta)\rVert.
\end{align}
Again, this optimization problem can be addressed directly using solvers for nonlinear programming.

\subsection{Exact computation of $\gamma$ via combinatorial optimization}\label{app:robustness_bounds_3}
In the following, we show that~\eqref{eq:gamma_bound_NL} can be reformulated equivalently as a combinatorial optimization problem.
To this end, note that~\eqref{eq:gamma_bound_NL} is equivalent to 
\begin{align}\label{eq:averaged_interaction_Hamiltonian_SDP_problem2}
    \gamma^2=\min_{\tilde{\gamma}\geq0}\>\>&\tilde{\gamma}\\\nonumber 
    \text{s.t.}\>\>&G(\theta)^\dagger G(\theta)\preceq \tilde{\gamma} I\>\>\forall\theta:\>\lVert\theta\rVert_{\infty}\leq \frac{1}{\ell},
\end{align}
where we write $A\preceq B$ if $A-B$ is negative semidefinite.
Since the constraint in~\eqref{eq:averaged_interaction_Hamiltonian_SDP_problem2} is convex in $\theta$, it suffices to enforce it only on the vertices of $\lVert\theta\rVert_{\infty}\leq \frac{1}{\ell}$, i.e.,~\eqref{eq:averaged_interaction_Hamiltonian_SDP_problem2} is equivalent to
\begin{align}
    \gamma^2=\min_{\tilde{\gamma}\geq0}\>\>&\tilde{\gamma}\\\nonumber 
    \text{s.t.}\>\>&G(\theta)^\dagger G(\theta)\preceq \tilde{\gamma}I\>\>\forall\theta\in\Big\{-\frac{1}{\ell},\frac{1}{\ell}\Big\}^{\ell N}.
\end{align}
This problem can, in turn, be reformulated as 
\begin{align}\label{eq:gamma_bound_combinatorial}
    \gamma=\sqrt{\tilde{\gamma}^*}=\max_{\theta\in\{-\frac{1}{\ell},\frac{1}{\ell}\}^{\ell N}}\sigma_{\max}(G(\theta)),
\end{align}
where $\sigma_{\max}(G(\theta))$ is the maximum singular value of $G(\theta)$.
Thus, a bound $\gamma$ can be obtained by computing the maximum singular value of $G(\theta)$ on the vertices $\{-\frac{1}{\ell},\frac{1}{\ell}\}^{\ell N}$ of the hyperrectangle $\lVert\theta\rVert_{\infty}\leq\frac{1}{\ell}$.

\subsection{Norm-based bound on $\gamma$}\label{app:robustness_bounds_4}
We rewrite $G(\theta)$ as
\begin{align}\label{eq:averaged_interaction_Hamiltonian_decomposition}
    \mathrm{vec}(G(\theta))=M\theta
\end{align}
with
\begin{align}
    M&=\frac{1}{N}\begin{bmatrix}
        M_1&\dots&M_N
    \end{bmatrix},\\\nonumber
    M_j&=\begin{bmatrix}
        \mathrm{vec}(\bar{V}_j^\dagger B_{j,1}\bar{V}_j)&\dots&
        \mathrm{vec}(\bar{V}_j^\dagger B_{j,{\ell}}\bar{V}_j)
    \end{bmatrix},\\\nonumber 
    \theta&=(\theta_1,\dots,\theta_N).
\end{align}
Note that
\begin{align}\label{eq:norm_based_gamma}
    &\lVert G(\theta)\rVert\leq\lVert G(\theta)\rVert_F=\lVert\mathrm{vec}(G(\theta))\rVert\\\nonumber
    \leq&\lVert M\rVert\lVert\theta\rVert\leq\lVert M\rVert\lVert\theta\rVert_{\infty}\sqrt{\ell N}
    \leq\sqrt{\frac{N}{\ell}}\lVert M\rVert\delta.
\end{align}
Thus, we obtain the upper bound 
\begin{align}\label{eq:gamma_bound_norm}
    \gamma\leq\sqrt{\frac{N}{\ell}}\lVert M\rVert.
\end{align}

\subsection{Discussion and comparison}
The computational procedures outlined above are complementary with different benefits and drawbacks.
Computing the bound via~\eqref{eq:F_bound_NL} or~\eqref{eq:gamma_bound_NL} is simple to implement and scales moderately.
However, when solving these optimization problems directly using a nonlinear programming solver, there is no guarantee that the obtained solution is globally optimal.
If it is not, then the estimated worst-case fidelity may in principle be larger than the true one defined in~\eqref{eq:thm_avg_algo}.
On the other hand, computing $\gamma$ via~\eqref{eq:gamma_bound_combinatorial} is guaranteed to return the true worst-case value of $\gamma$.
The downside is that~\eqref{eq:gamma_bound_combinatorial} contains $2^{\ell N}$ constraints and, therefore, it can only be implemented for very small numbers of gates and qubits.
Finally, computing an upper bound on $\gamma$ as in~\eqref{eq:gamma_bound_norm} is simple to implement and scalable, only requiring to compute a matrix norm, but it is in general conservative.
In particular, it may lead to a loose upper bound on $\gamma$ and, therefore, to an unnecessarily small worst-case fidelity bound in~\eqref{eq:thm_avg_algo}.

It is important to emphasize that none of these approaches scale directly to high numbers of qubits or gates.
In particular, each method requires to classically evaluate the matrix $G(\theta)$, whose size grows exponentially in the number of qubits.
Moreover, the computation~\eqref{eq:gamma_bound_combinatorial} additionally scales exponentially with the number of gates.
In Appendix~\ref{app:circuit_partitioning}, we address this problem via a circuit partitioning strategy, which allows us to compute bounds on $\gamma$ in a scalable fashion, paving the way for relevant problem sizes beyond classical simulation.

\section{Scalable bounds on $\gamma$ via circuit partitioning}\label{app:circuit_partitioning}

Computing the bounds in Appendix~\ref{app:robustness_bounds} is intractable for circuits which cannot be classically simulated.
As we show in the following, bounds on $\lVert G\rVert$ can be computed in a modular and parallelizable fashion, allowing us to scale to larger circuits at the cost of a possible increase of conservatism.
In this way, we can scale up all approaches from Section~\ref{app:robustness_bounds} which rely on a bound on $\lVert G\rVert$, i.e., approaches B, C, and D.
The proposed procedure is related to circuit cutting techniques~\cite{peng2019simulating,lowe2023fast,vazquez2024scaling} with the main difference that bounding $\lVert G\rVert$ does not require keeping track of the intermediate quantum state.

In order to bound $\lVert G\rVert$, we partition the circuit $\bar{U}$ into two pieces 
\begin{align}
    \bar{U}&=\underbrace{\bar{U}_N\cdots \bar{U}_{n_\rmA+1}}_{\bar{U}^\rmB}
    \underbrace{\bar{U}_{n_\rmA}\cdots\bar{U}_1}_{\bar{U}^\rmA}
\end{align}
such that $\bar{U}^\rmA$ can be efficiently evaluated classically (e.g., it only acts on a small subset of qubits).
A bound on $\lVert G\rVert$ can then be computed sequentially as
\begin{align}\label{eq:scalability_sequential_bound}
    \lVert G\rVert&=\Big\lVert\frac{1}{N}\sum_{j=1}^NG_j\Big\rVert\\\nonumber 
    &\leq\frac{1}{N}\Big(\Big\lVert\sum_{j=1}^{n_\rmA}G_j\Big\rVert+
    \Big\lVert\sum_{j=n_\rmA+1}^{N}G_j\Big\rVert\Big).
\end{align}
In the following, we discuss how the two terms on the right-hand side of~\eqref{eq:scalability_sequential_bound} can be computed. 
Regarding the first term, if $\bar{U}^\rmA$ can be efficiently evaluated classically, then we can bound 
\begin{align}
    \Big\lVert\sum_{j=1}^{n_\rmA}G_j\Big\rVert
\end{align}
as in Appendix~\ref{app:robustness_bounds}.
On the other hand, the second term on the right-hand side of~\eqref{eq:scalability_sequential_bound} is equal to ($N-n_{\rmA}$ times) the norm of the averaged interaction Hamiltonian $\braket{G^\rmB}$ for the circuit 
\begin{align}
    U'=U_N\cdots U_{n_\rmA+1},
\end{align}
which can also be computed as in Section~\ref{app:robustness_bounds}.
To show this, we use the unitarity of the involved gates to obtain
\begin{align}\label{eq:averaged_interaction_hamiltonian_scalable_bound}
    &\quad\Big\lVert\sum_{j=n_\rmA+1}^{N}G_j\Big\rVert=
    \Big\lVert\sum_{j=n_\rmA+1}^{N}\theta_j\bar{V}_j^\dagger H_{\rme,j}\bar{V}_j\Big\rVert\\\nonumber 
    &=\Big\lVert\sum_{j=n_\rmA+1}^{N}
    \theta_j(\bar{U}_{j-1}\cdots\bar{U}_1)^\dagger H_{\rme,j}
    \bar{U}_{j-1}\cdots\bar{U}_1
    \Big\rVert\\\nonumber 
    &=\Big\lVert
        \bar{V}_{n_{\rmA}+1}^\dagger
        \Big(
        \theta_{n_{\rmA}+1}H_{\rme,n_{\rmA}+1}+\\\nonumber 
        &
        \sum_{j=n_\rmA+2}^N 
        (\bar{U}_{j-1}\cdots\bar{U}_{n_\rmA+1})^\dagger H_{\rme,j}
        \bar{U}_{j-1}\cdots\bar{U}_{n_\rmA+1}\Big)\bar{V}_{n_{\rmA}+1}
        \Big\rVert\\\nonumber
        &=\Big\lVert
            \theta_{n_{\rmA}+1}H_{\rme,n_{\rmA}+1}+\\\nonumber 
            &\sum_{j=n_\rmA+2}^N \theta_j
            (\bar{U}_{j-1}\cdots\bar{U}_{n_\rmA+1})^\dagger H_{\rme,j}
            \bar{U}_{j-1}\cdots\bar{U}_{n_\rmA+1}
        \Big\rVert.
\end{align}
This expression corresponds to the norm of the averaged interaction Hamiltonian $\braket{G^\rmB}$ corresponding to the reduced circuit $\bar{U}^\rmB$.
Thus, if $\bar{U}^\rmB$ can be simulated classically, then we can compute the bound~\eqref{eq:averaged_interaction_hamiltonian_scalable_bound} as in Section~\ref{app:robustness_bounds}.
On the other hand, if $\bar{U}^\rmB$ cannot be simulated classically, then we can proceed recursively by cutting it into smaller pieces.
Notably, bounds for the individual pieces can be obtained independently, which allows us to parallelize the computations. 

\section{Comparison to existing works}\label{app:comparison}

\subsection{Special case: coherent control errors}\label{app:comparison_coherent_control}
In the following, we compare the worst-case fidelity bound from Theorem~\ref{thm:avg_algo} to~\cite{berberich2024robustness}, which provides a worst-case fidelity bound for the special case of coherent control errors, compare~\eqref{eq:application_noisy_circuit}.
The main fidelity bound from~\cite[Theorem 2]{berberich2024robustness} states 
\begin{align}\label{eq:fidelity_bound_pra}
    \sqrt{F_{\mathrm{wc}}}\geq1-\Big(\sum_{j=1}^N\lVert \bar{H}_j\rVert\Big)^2\frac{\lVert\theta\rVert_{\infty}^2}{2}.
\end{align}
Using $a\geq2\sqrt{a}-1$ for arbitrary $a>0$, this fidelity bound can be reformulated as 
\begin{align}\label{eq:fidelity_bound_pra_2}
    F_{\mathrm{wc}}\geq2\sqrt{F_{\mathrm{wc}}}-1\geq 1-\Big(\sum_{j=1}^N\lVert \bar{H}_j\rVert\Big)^2\lVert\theta\rVert_{\infty}^2.
\end{align}
Note that $\sum_{j=1}^N\lVert H_j\rVert$ scales proportionally to $N$, and proportionally to $nN$ when $n$ gates acting on different qubits are implemented in each layer.
In case of only one gate per layer,~\eqref{eq:fidelity_bound_pra_2} implies
\begin{align}
    F_{\mathrm{wc}}\geq 1-c\delta^2N^2
\end{align}
for some $c>0$.
Hence, to guarantee a given worst-case fidelity $\bar{F}_{\mathrm{wc}}$, we need that 
\begin{align}\label{eq:scaling_infidelity_bound_pra}
    c\delta^2N^2\leq1-\bar{F}_{\mathrm{wc}}.
\end{align}
Note the difference between~\eqref{eq:scaling_infidelity_bound} and~\eqref{eq:scaling_infidelity_bound_pra}.
When $\delta N$ is small compared to $\gamma$, both infidelity bounds scale as $\delta^2N^2$.
However, for robust algorithms with $\gamma\approx0$, the bound~\eqref{eq:scaling_infidelity_bound} derived in the present paper scales in a significantly more favorable way with $\delta^4N^4$.
In Section~\ref{subsec:numerical_results_circuit_transpilation}, we compare both bounds for a concrete set of algorithms, showing that the bound obtained via Theorem~\ref{thm:avg_algo} is substantially less conservative than the bound by~\cite{berberich2024robustness}.

\subsection{Stochastic analysis from~\cite{garcia2025resilience}}
Next, we compare Theorem~\ref{thm:avg_algo} to \cite{garcia2025resilience}.
This work considers the infidelity measure $F_{Q}=2(1-\braket{\psi|\hat{\psi}})$ which provides a lower bound on the fidelity~\eqref{eq:fidelity} as 
\begin{align}
    F\geq1-F_Q.
\end{align}
Contrary to our worst-case analysis,~\cite{garcia2025resilience} takes a stochastic viewpoint.
Assuming uncorrelated and normally distributed noise, and after an averaging step, it is shown that
\begin{align}\label{eq:app_comparison_F_Q_average}
    F_Q\sim Nn\delta^2.
\end{align}
This bound indicates a favorable scaling of the error level with the circuit depth, i.e., $\delta\sim\frac{1}{\sqrt{N}}$ is sufficient for high fidelity, whereas the bound in Theorem~\ref{thm:avg_algo} results in $\delta\sim\frac{1}{N}$.
However, it is important to emphasize that the bound~\eqref{eq:app_comparison_F_Q_average} is only shown in~\cite{garcia2025resilience} for the average fidelity and under independent and normally distributed noise.
On the contrary, our results address arbitrary (bounded) noise and worst-case fidelity bounds, which are commonly assumed for threshold theorems~\cite{nielsen2011quantum,kitaev1997quantum,aharonov1997fault,aharonov2008fault}.
In particular, a scaling of $\delta\sim\frac{1}{N}$ can be expected for a general worst-case bound against coherent errors and without further structural assumptions since it cannot be excluded that the errors act in the same direction, which causes them to accumulate linearly.

We provide a simple example to demonstrate this fact.
Consider a trivial algorithm $\bar{U}=I$ and single-qubit error unitaries $U_{e,j}=e^{-i\varepsilon Z}$ with identical values $\varepsilon\in\bbr$ for the different errors $j=1,\dots,N$.
The resulting noisy algorithm can then be expressed as $U=e^{-i\varepsilon NZ}$.
Assuming that $\lVert\varepsilon\rVert\leq\delta$ for some $\delta>0$, it is straightforward to show that the fidelity is equal to 
\begin{align}
    F(\bar{U},U)&=\Big|\frac{1}{2}\rmtr(e^{-i\varepsilon NZ})\Big|^2\\\nonumber 
    &=\Big|\frac{1}{2}(e^{-i\varepsilon N}+e^{i\varepsilon N})\Big|^2=|\cos(\delta N)|^2\\\nonumber 
    &=1-\delta^2N^2+\calO(\delta^4N^4).
\end{align}
This proves that, indeed, for small $\delta N$, the infidelity scales with $\delta^2N^2$ in the worst case.
Hence, to keep the same level of fidelity for increasing $N$, the noise level needs to scale as $\delta\sim\frac{1}{N}$, which matches our theoretical lower bound.

\section{Proof of Theorem~\ref{thm:avg_operations}}\label{app:proof_thm_avg_operations}

\begin{proof} 
\textbf{Part (i): Proof of~\eqref{eq:thm_avg_operations_aux}}\\
We can bound $\lVert \calR_N-I\rVert$ as
\begin{align}\label{eq:thm_avg_operations_proof0}
    &\lVert\calR_N-I\rVert\leq\lVert\calR_N-\bar{\calR}_N\rVert 
    +\lVert\bar{\calR}_N-I\rVert,
\end{align}
where $\bar{\calR}_N=e^{N\calG}$.
The first term on the right-hand side can be bounded in complete analogy to the proof of Theorem~\ref{thm:avg_algo}
\begin{align}\label{eq:thm_avg_operations_proof_R_bound}
    \lVert\calR_N-\bar{\calR}_N\rVert \leq 
    \frac{1}{2}\sum_{j=1}^N\Big\lVert \sum_{k=j+1}^N [\calG_j,\calG_k]\Big\rVert,
\end{align}
compare~\eqref{eq:thm_avg_algo_proof_2}.
Moreover, to bound the second term on the right-hand side of~\eqref{eq:thm_avg_operations_proof0}, we define 
\begin{align}\label{eq:thm_avg_operations_proof_f_def}
    f(x)=e^{xN\calG}.  
\end{align}
We compute a Lipschitz bound of $f$ via the norm of its derivative
\begin{align}
    \Big\lVert\frac{\rmd}{\rmd x}f(x)\Big\rVert=\lVert N\calG e^{xN\calG}\rVert.
\end{align}
We write $\calL_{\calG}$ for the Lindblad superoperator $\rho\mapsto\calL_{\calG}(\rho)$ which corresponds to the differential equation $\dot{x}=\calG x$, where $x=\mathrm{vec}(\rho)$ is the vectorized state.
Since quantum operations do not increase the trace norm, we have
\begin{align}\label{eq:thm_avg_operations_proof_contraction}
    \rmtr|\calL_{\calG}(\rho)|\leq\rmtr|\rho|.
\end{align}
We now use the equivalence of the trace norm and the Frobenius norm
\begin{align}\label{eq:thm_avg_operations_proof_norm_equivalence}
    \lVert\rho\rVert_F\leq\rmtr|\rho|\leq2^{n/2}\lVert\rho\rVert_F.
\end{align} 
This implies 
\begin{align}\label{eq:thm_avg_operations_proof_trace_preserving}
    \lVert \calL_{\calG}(\rho)\rVert_F\stackrel{\eqref{eq:thm_avg_operations_proof_norm_equivalence}}{\leq}
\rmtr|\calL_{\calG}(\rho)|\stackrel{\eqref{eq:thm_avg_operations_proof_contraction}}{\leq} 
\rmtr|\rho|\stackrel{\eqref{eq:thm_avg_operations_proof_norm_equivalence}}{\leq}
2^{n/2}\lVert\rho\rVert_F.
\end{align}
Using $\lVert\rho\rVert_F=\lVert\mathrm{vec}(\rho)\rVert$, this means that 
\begin{align}
    \lVert \mathrm{vec}(\calL_{\calG}(\rho))\rVert\leq&2^{n/2}\lVert\mathrm{vec}(\rho)\rVert.
\end{align}
Note that $\mathrm{vec}(\calL_{\calG}(\rho))=e^{t\calG}\mathrm{vec}(\rho)$ for some $t\geq0$.
Thus, we have shown that the induced $2$-norm of $e^{xN\calG}$ is bounded as $\lVert e^{xN\calG}\rVert\leq 2^{n/2}$ and, therefore, a Lipschitz bound of the map $f$ defined in~\eqref{eq:thm_avg_operations_proof_f_def} is given by $2^{n/2}N\lVert\calG\rVert$.
This implies 
\begin{align}\label{eq:thm_avg_operations_proof_R_bar_bound}
    \lVert\bar{\calR}_N-I\rVert=\lVert f(1)-f(0)\rVert\leq 2^{n/2}N\lVert\calG\rVert.
\end{align}
Combining~\eqref{eq:thm_avg_operations_proof0}, \eqref{eq:thm_avg_operations_proof_R_bound}, and \eqref{eq:thm_avg_operations_proof_R_bar_bound} we obtain 
\begin{align}\label{eq:thm_avg_operations_proof_R_identity_bound}
    \lVert\calR_N-I\rVert\leq \frac{1}{2}\sum_{j=1}^N\Big\lVert \sum_{k=j+1}^N [\calG_j,\calG_k]\Big\rVert+2^{n/2}N\lVert\calG\rVert.
\end{align}
Next, we relate this bound to the minimal fidelity defined in~\eqref{eq:F_min_def}.
Recall that
\begin{align}\label{eq:thm_avg_operations_proof1}
    \calF_{\mathrm{min}}(\bar{\calE},\calE)=& \min_{\rho\>\text{pure}}\calF_{\mathrm{state}}
    ( \bar{\calE}(\rho),
    \calE(\rho)).
\end{align}
For the trace distance $D_{\mathrm{state}}(\rho,\sigma)=\frac{1}{2}\rmtr|\rho-\sigma|$, it holds that 
\begin{align}
    \sqrt{\calF_{\mathrm{state}}(\rho,\sigma)}\geq 1-D_{\mathrm{state}}(\rho,\sigma),
\end{align}
see~\cite[equation (5)]{gilchrist2005distance}.
Inserting this into~\eqref{eq:thm_avg_operations_proof1}, we infer 
\begin{align}
    &\sqrt{\calF_{\mathrm{min}}(\bar{\calE},\calE)}
    \geq
    1-\max_{\rho\>\text{pure}}D_{\mathrm{state}}( \bar{\calE}(\rho),
    \calE(\rho)).
\end{align}
Thus, using~\eqref{eq:thm_avg_operations_proof_norm_equivalence}, we have
\begin{align}\label{eq:thm_avg_operations_proof2}
    \sqrt{\calF_{\mathrm{min}}(\bar{\calE},\calE)} 
    \geq&1-2^{n/2-1}\max_{\rho\>\text{pure}}\lVert \bar{\calE}(\rho)-\calE(\rho)\rVert_F\\\nonumber 
    =&1-2^{n/2-1}\max_{\rho\>\text{pure}}\lVert (\bar{\calA}-\calA)\mathrm{vec}(\rho)\rVert\\\nonumber 
    \geq &1-2^{n/2-1}\lVert \bar{\calA}-\calA\rVert,
\end{align}
where we use that the matrix $2$-norm is induced by the vector $2$-norm in the last inequality.
Using straightforward algebraic manipulations analogous to the setup in Section~\ref{sec:robustness_quantum_algorithms}, one can prove that $\calA=\bar{\calA}\calR_N$.
Hence, the deviation of the noisy operation $\calA$ from the ideal operation $\bar{\calA}$ can be quantified by the deviation of $\calR_N$ from identity.
In particular, we have
\begin{align}\label{eq:thm_avg_operations_proof3}
    \sqrt{\calF_{\mathrm{min}}(\bar{\calE},\calE)}\geq&1-2^{n/2-1}
    \lVert (I-\calR_N)\bar{\calA}\rVert\\\nonumber 
    \geq&1-2^{n/2-1}
    \lVert I-\calR_N\rVert\lVert\bar{\calA}\rVert.
\end{align}
Analogously to~\eqref{eq:thm_avg_operations_proof_trace_preserving}, one can show that $\lVert\bar{\calA}\rVert\leq2^{n/2}$.
Together with~\eqref{eq:thm_avg_operations_proof_R_identity_bound} and $\Big(1-\sqrt{\calF_{\min}(\bar{\calE},\calE)}\Big)^2\geq0$, this leads to
\begin{align}\label{eq:thm_avg_operations_proof4}
    &\calF_{\min}(\bar{\calE},\calE)\geq2\sqrt{\calF_{\min}(\bar{\calE},\calE)}-1\\\nonumber 
    \geq&1-2^{n}
    \left(\frac{1}{2}\sum_{j=1}^N\Big\lVert\sum_{k=j+1}^N[\calG_j,\calG_k]\Big\rVert 
    +2^{n/2}N\lVert\calG\rVert\right),
\end{align}
which proves~\eqref{eq:thm_avg_operations_aux}.\\
\textbf{Part (ii): Proof of~\eqref{eq:thm_avg_operations}}\\
First, note that for unitary operations $\bar{\calE}$, it holds that $\lVert\bar{\calA}\rVert=1$ and, therefore,~\eqref{eq:thm_avg_operations_proof4} can be tightened to 
\begin{align}\label{eq:thm_avg_operations_proof5}
    &\calF_{\min}(\bar{\calE},\calE)\\\nonumber 
    \geq&1-2^{n/2}
    \left(\frac{1}{2}\sum_{j=1}^N\Big\lVert\sum_{k=j+1}^N[\calG_j,\calG_k]\Big\rVert 
    +2^{n/2}N\lVert\calG\rVert\right).
\end{align}
Similar to the proof of Theorem~\ref{thm:avg_algo}, we have 
\begin{align}\label{eq:thm_avg_operations_proof6}
    \frac{1}{2}\sum_{j=1}^N\Big\lVert\sum_{k=j+1}^N[\calG_j,\calG_k]\Big\rVert 
    \leq\frac{N^2-N}{2}\max_{j=1,\dots,N}\lVert\calG_j\rVert^2.
\end{align}
Using that the $\bar{\calA}_j$'s are unitary, we have 
\begin{align}
\lVert\calG_j\rVert=\lVert M_{\rme,j}\rVert
\stackrel{\eqref{eq:thm_avg_operations_ass1}}{\leq}\delta.
\end{align}
Combining this with~\eqref{eq:thm_avg_operations_ass2}, \eqref{eq:thm_avg_operations_proof5}, and~\eqref{eq:thm_avg_operations_proof6}, we obtain 
\begin{align}
    \calF_{\min,\mathrm{wc}}\geq
    1-2^{n/2}\left( \frac{N^2-N}{2}\delta^2+2^{n/2}N\gamma\delta\right),
\end{align}
which proves~\eqref{eq:thm_avg_operations}.
\end{proof}

\section{Details on the numerical experiments}\label{app:details_numerics}
\subsection{QFT Circuits}\label{app:details_numerics_qft_circuits}
\begin{figure}
    \centering
    \includegraphics[width=\linewidth]{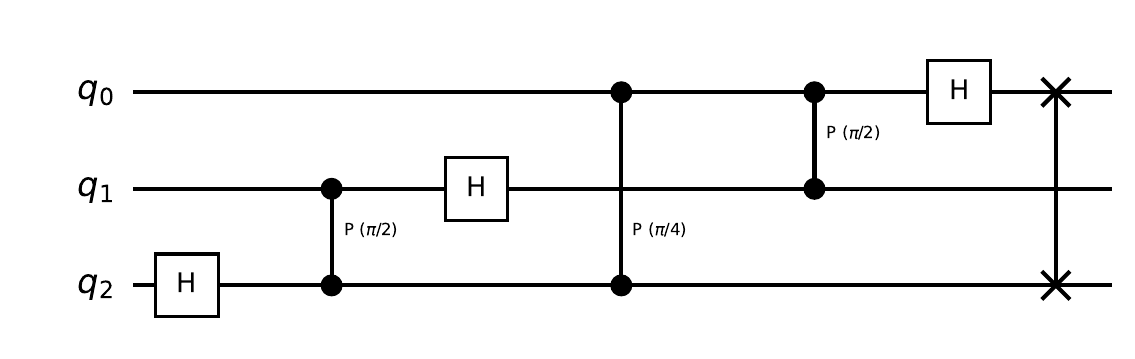}
    \caption{Circuit diagram of the untranspiled three-qubit textbook QFT circuit.}
    \label{fig:qft_circuit_original}
\end{figure}

\begin{figure}
    \centering
    \includegraphics[width=\linewidth]{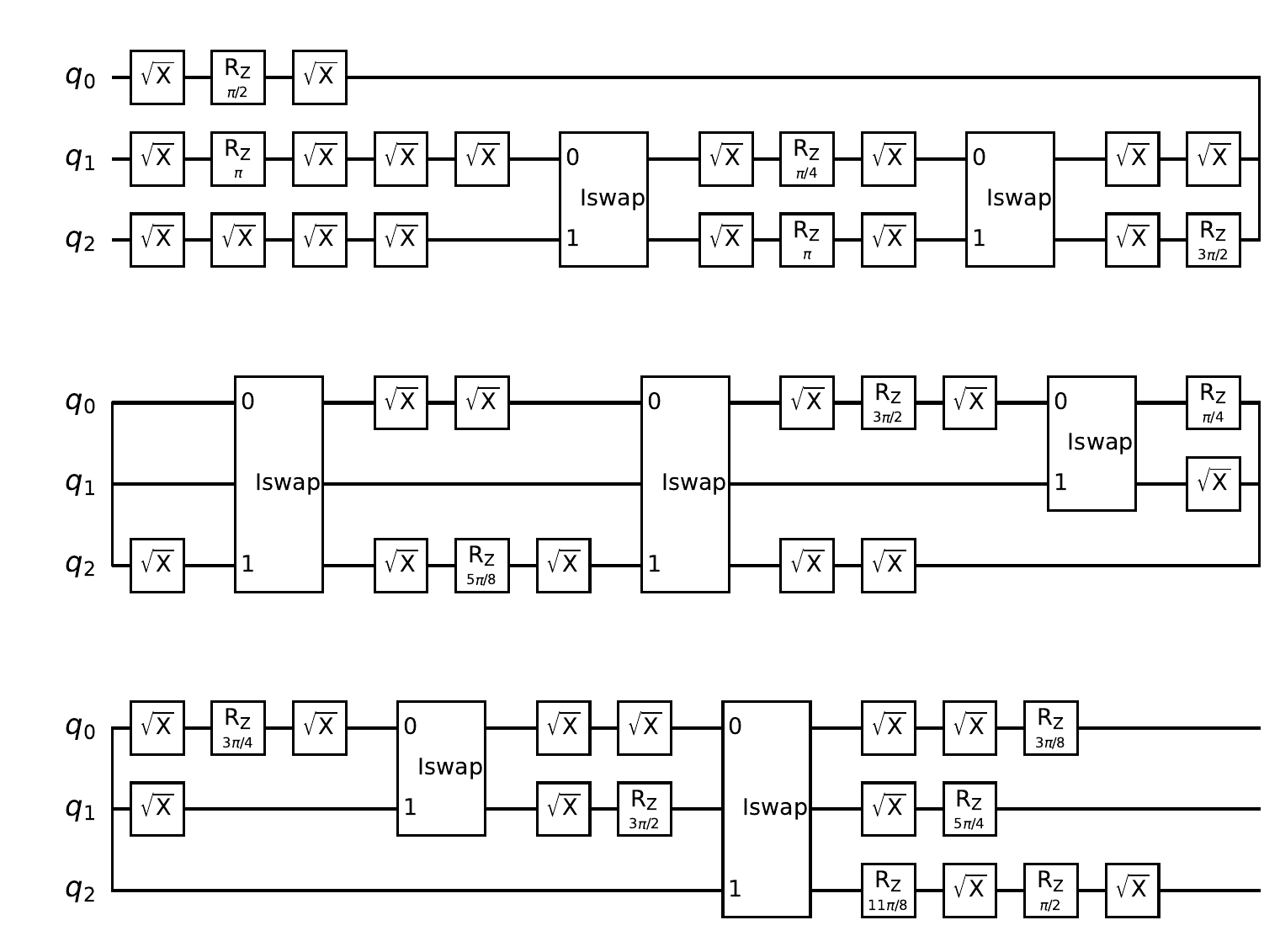}
    \caption{Transpiled quantum circuit for the three-qubit QFT with gate set A.}
    \label{fig:qft_circuit_rigetti}
\end{figure}

\begin{figure}
    \centering
    \includegraphics[width=\linewidth]{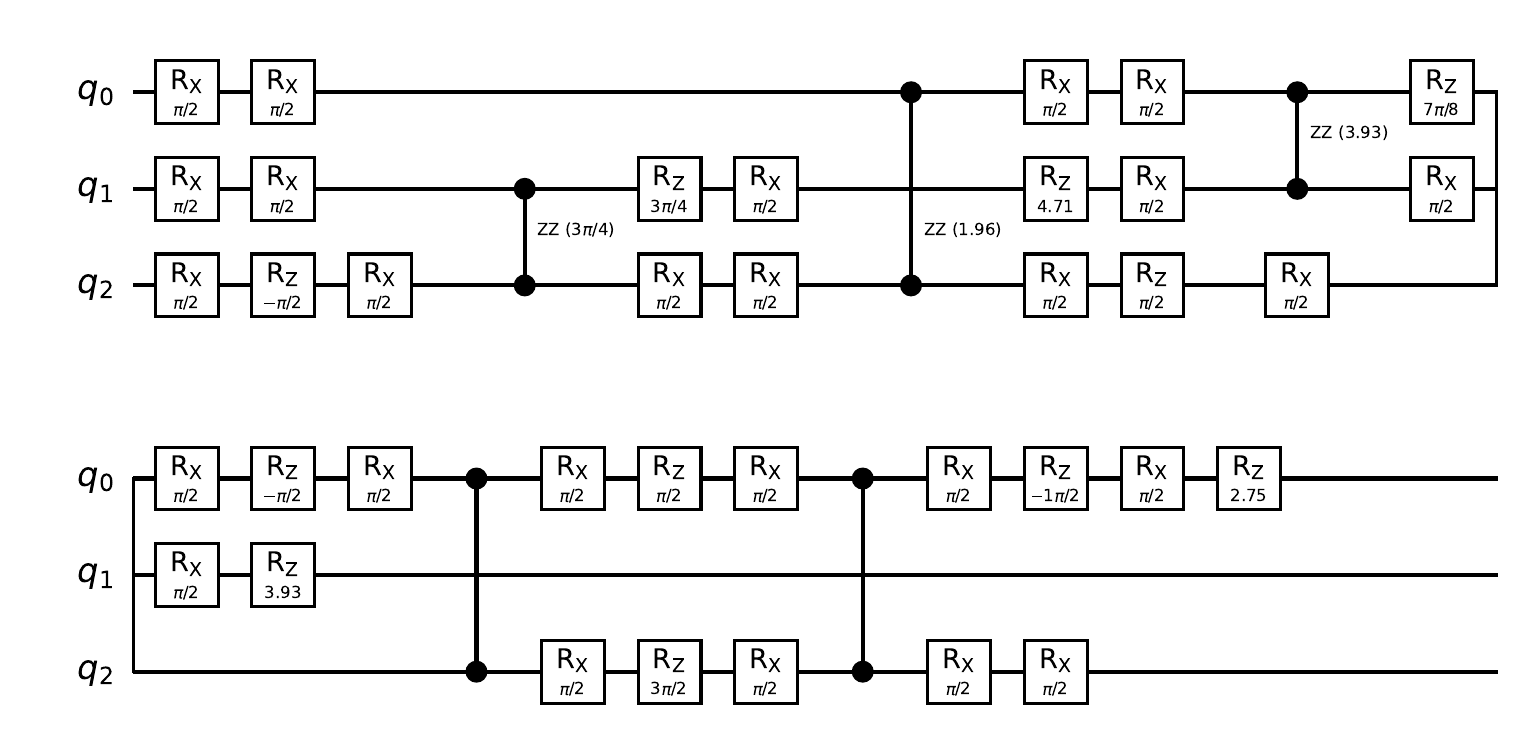}
    \caption{Transpiled quantum circuit for the three-qubit QFT with gate set B.}
    \label{fig:qft_circuit_ibm}
\end{figure}

\begin{figure}
    \centering
    \includegraphics[width=\linewidth]{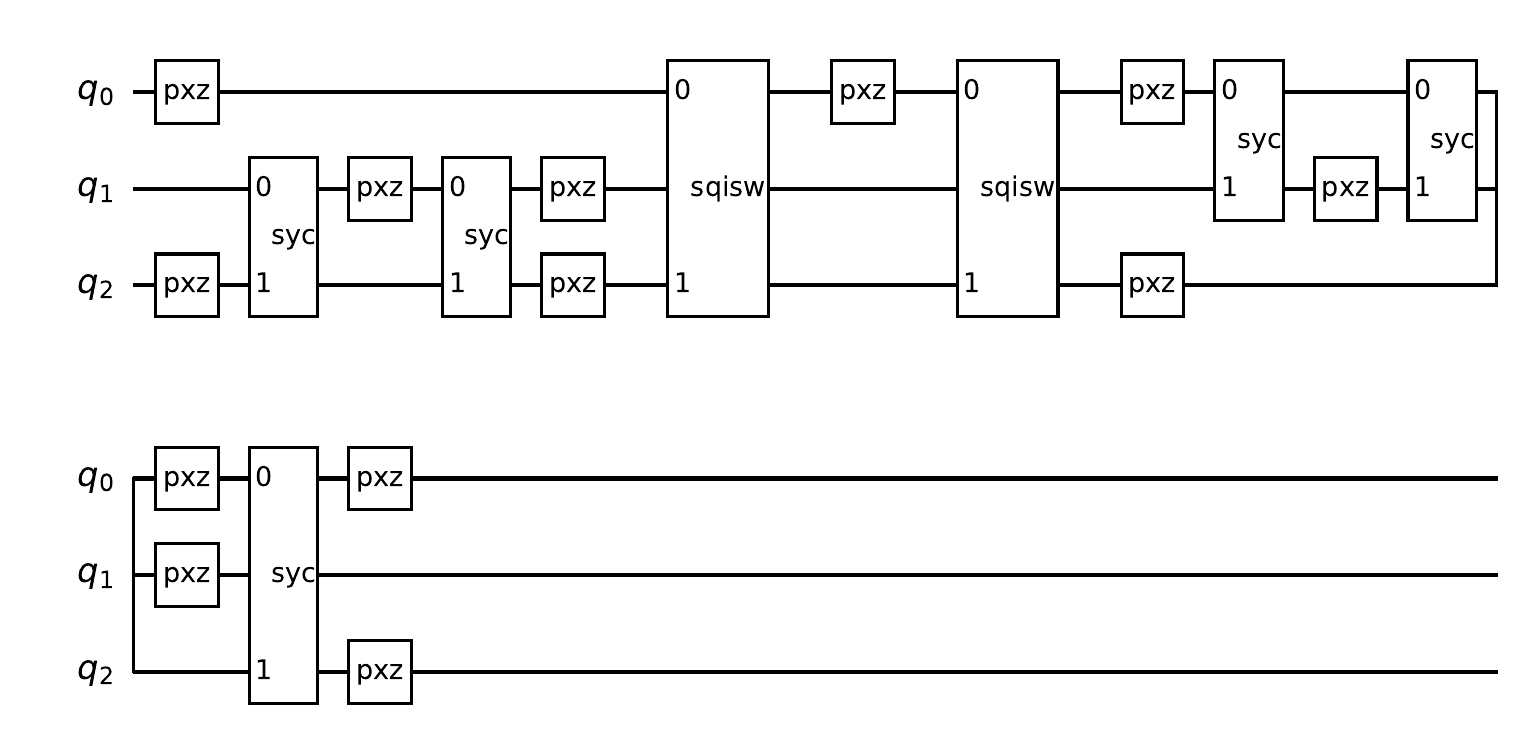}
    \caption{Transpiled quantum circuit for the three-qubit QFT with gate set C.}
    \label{fig:qft_circuit_google}
\end{figure}

\begin{figure}
    \centering
    \includegraphics[width=\linewidth]{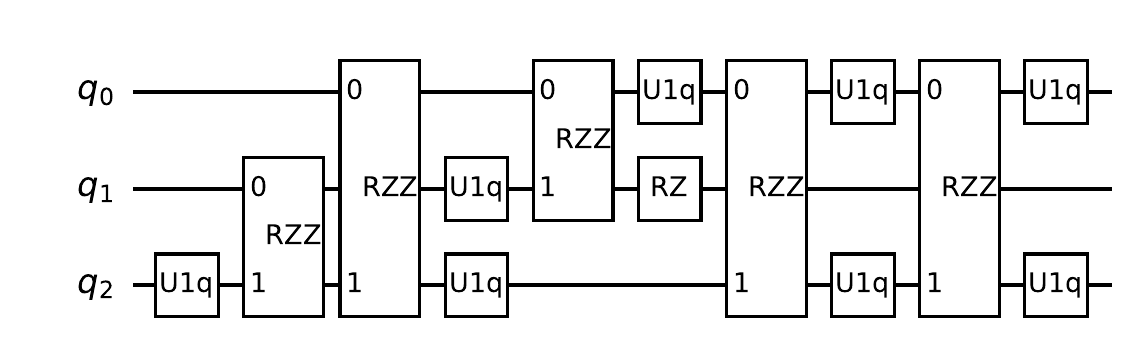}
    \caption{Transpiled quantum circuit for the three-qubit QFT with gate set D.}
    \label{fig:qft_circuit_quantinuum}
\end{figure}

Figure~\ref{fig:qft_circuit_original} shows the circuit diagram of untranspiled three-qubit textbook QFT circuit. We transpile the three-qubit textbook circuit into elementary base sets using the Berkeley Quantum Synthesis Toolkit (BQSKit)~\cite{bqskit_software} with optimization level 2. The transpiled circuits for gate sets A-D results in the circuits shown in Figures~\ref{fig:qft_circuit_rigetti}-\ref{fig:qft_circuit_quantinuum}.


\subsection{Modular adder circuit}\label{app:details_numerics_adder_circuit}
The modular adder circuit used in the numerical experiment of this paper is based on~\cite{gidney_factoring_2018}. We use the \texttt{qiskit}~\cite{javadi-abhariQuantumComputingQiskit2024} implementation of this circuit and transpile this circuit into gate set~B with the \texttt{transpile} method in \texttt{qiskit}.

\subsection{Details on numerically finding $\gamma$ via Appendix~\ref{app:robustness_bounds_2}}\label{app:details_numerics_gamma_opt}
To determine the parameter $\gamma$ via the reformulation as an optimization problem (Appendix~\ref{app:robustness_bounds_2}), we employ the L-BFGS-B algorithm~\cite{liu1989limited} as implemented in \texttt{scipy}. The optimization is initialized from 100 randomly chosen starting points, and we select the largest $\gamma$ value obtained across all runs for further analysis. We verified that the algorithm converges reliably in all cases. Although we cannot guarantee that the global maximum is always found, the observed variation between the largest values obtained is marginal, providing strong evidence that the resulting $\gamma$ values are close to the global optimum.

\section{Effect of the partition size $\gamma$}\label{app:partition_scaling}
The tightness of the bound on $\gamma$ is highly sensitive to the partition size and the specific bounding method employed. Generally, increasing the partition size (i.e., the maximum number of qubits in each sub-circuit) reduces the total number of partitions, thereby decreasing the conservatism introduced by the sequential bounding in~\eqref{eq:scalability_sequential_bound}. However, this comes at the cost of computational complexity, which scales exponentially with the number of qubits.

\begin{figure}[ht]
    \centering
    \includegraphics[width=\linewidth]{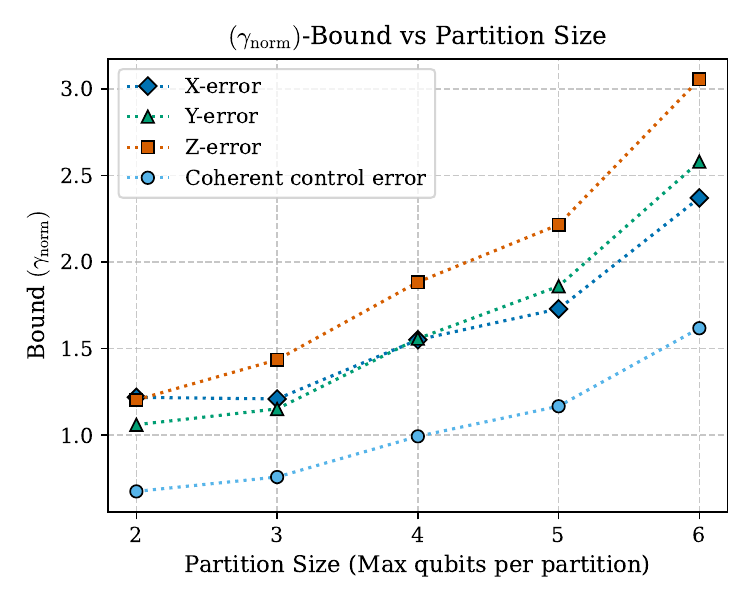}
    \caption{The bound on $\gamma_\mathrm{norm}$ for a 50 qubit modular adder where the circuit is partitioned into smaller sub-circuits with different partition sizes.}
    \label{fig:gamma_bound_vs_partition_size}
\end{figure}

When calculating the norm-based bound $\gamma_\mathrm{norm}$ (Appendix~\ref{app:robustness_bounds_4}), an additional competing effect arises from the Frobenius norm used in the inequalities of~\eqref{eq:norm_based_gamma}. This causes the bound to scale with the number of qubits involved in the circuit. By maintaining small partition sizes, the conservatism introduced by these norm inequalities remains manageable relative to that of the partition-based summation. This trade-off is illustrated in Figure~\ref{fig:gamma_bound_vs_partition_size}, which depicts the $\gamma_\mathrm{norm}$ bound for the 50-qubit modular adder across various partition sizes. The data demonstrates that for this specific method, smaller partition sizes can lead to less conservative estimates.

\end{document}